\documentclass[a4paper,11pt]{article}
\pdfoutput=1 

\usepackage{jheppub} 

\usepackage[T1]{fontenc} 

\usepackage{braket}

\title{\boldmath Quantum Nature of Black Holes: \\ Fast Scrambling versus Echoes}

\author[a,b,c]{Krishan Saraswat}
\author[a,b,c]{and Niayesh Afshordi}

\affiliation[a]{Department of Physics and Astronomy, University of Waterloo, 200 University Ave W, Waterloo, Canada}
\affiliation[b]{Waterloo Centre for Astrophysics, University of Waterloo, Waterloo, ON, N2L 3G1, Canada}
\affiliation[c]{Perimeter Institute For Theoretical Physics, 31 Caroline St N, Waterloo, Canada}

\emailAdd{ksaraswat@pitp.ca}
\emailAdd{nafshordi@pitp.ca}

\abstract{Two seemingly distinct notions regarding black holes have captured the imagination of theoretical physicists over the past decade: First, black holes are conjectured to be fast scramblers of information, a notion that is further supported through connections to quantum chaos and decay of mutual information via AdS/CFT holography. Second, black hole information paradox has motivated exotic quantum structure near horizons of black holes (e.g., gravastars,  fuzzballs, or firewalls) that may manifest themselves through delayed gravitational wave echoes in the aftermath of black hole formation or mergers, and are potentially observable by LIGO/Virgo observatories. By studying various limits of charged AdS/Schwarzschild black holes we show that, if properly defined, the two seemingly distinct phenomena happen on an identical timescale of  log(Radius)/$(\pi \times {\rm Temperature})$. We further comment on the physical interpretation of this coincidence and the corresponding holographic interpretation of black hole echoes.   }

\begin{document} 
\maketitle

\section{Introduction}
\label{introsec}
Recent studies of black holes from the point of view of string theory and quantum information suggest that the horizon of a black hole may be modified. Most notably, modified horizons appear in the context of the black hole information paradox in the form of a firewall  \cite{Almheiri:2012rt,Polchinski:2016hrw,Stoica:2018uli} and also within string theory in the tight fuzzball paradigm \cite{Mathur:2009hf,Mathur:2013bra,Guo:2017jmi}. These descriptions usually suggest modifications within a Planck length of the horizon, we refer to these as ``hard'' modifications. These are in contrast to studies which suggest ``soft'' modifications which can manifest as soon as one gets within a black hole radius of the horizon \cite{Giddings:2017mym,Giddings:2019vvj,Buoninfante:2019swn}. Furthermore, recent experimental results from the detection of gravitational waves have provided tentative (albeit controversial) evidence of modified horizons \cite{Abedi:2016hgu,Conklin:2017lwb,Abedi:2018npz} (see \cite{Westerweck:2017hus} and \cite{Abedi:2018pst}, for counterpoint and rebuttal). A particularly interesting property of black holes with modified horizons comes from the study of its quasi-normal modes. Typically, quasi-normal modes of a black hole are found by requiring in-going boundary conditions at the horizon \cite{Horowitz:1999jd,Berti:2009kk}. However, for black holes with modified horizons, it is believed that such boundary conditions will be altered. One way to model such changes is to introduce boundary conditions on a surface which exists within a proper Planck length of the horizon\footnote{Since these modifications are localized within a Planck length of the horizon we would classify these as ``hard'' modifications.}. This surface or membrane allows for the partial reflection of perturbations. Studies using this approach have shown that the quasi-normal modes exhibit ``echoes''\cite{Cardoso:2016rao,Cardoso:2016oxy,Mark:2017dnq,Cardoso:2017cqb,Cardoso:2017njb,Wang:2018gin,Cardoso:2019rvt}. The term ``echoes'' is used to refer to a feature of the late time decay behaviour of the quasi-normal modes. For typical black holes (i.e. black holes with smooth horizons) the decay is exponential. For black holes with modified horizons the late time behaviour is accompanied by small repeating peaks in the amplitude. The physical reason why one sees repeating peaks in the amplitude is because perturbations will bounce back and fourth between the modified horizon and angular momentum barrier (similar to echoes created using sound waves). The time delay between adjacent peaks is referred to as the {\it echo time}. The echo time, in the geometric optics approximation, is twice the tortoise coordinate distance between the modified horizon/membrane and angular momentum barrier \cite{Abedi:2016hgu,Wang:2018gin}:
\begin{equation}
t_{echo} \simeq 2 |r_*|_{\rm membrane}.
\end{equation}
It was first noted in \cite{Abedi:2016hgu} that the echo time was comparable to the scrambling time scale for black holes. 

The scrambling time scale appears when black holes are studied from an information theoretic point of view \footnote{Usually these types of studies assume that black hole evaporation is unitary.}. In the context of quantum information recovery, the scrambling time scale can be viewed as a lower bound on the time it takes between throwing information into a black hole and being able to recover it, with small error from the subsequent Hawking radiation \cite{Hayden:2007cs,Sekino:2008he,Yoshida:2017non}. It has also been described as the amount of time it takes for a qubit of information thrown into a black hole to become thoroughly ``mixed'' \cite{Sekino:2008he,Lashkari:2011yi}. There are many methodologies in the current literature to calculate the scrambling time scale for black holes \cite{Sekino:2008he,Lashkari:2011yi,Shenker:2013yza,Leichenauer:2014nxa,Brown:2018kvn}. Depending on the particular approach one takes the exact mathematical expression for the scrambling time scale may vary. However, as diverse as they may be, it seems that the approaches described in \cite{Sekino:2008he,Lashkari:2011yi,Shenker:2013yza,Dvali:2013vxa,Leichenauer:2014nxa,Brown:2018kvn} give a time scale that can be roughly quantified by the following expression\footnote{This is not to say that every approach to compute scrambling time gives a time scale similar to Eq. (\ref{GenericScr}). A notable exception is suggested by Peter Shor in \cite{Shor:2018sws}, which we will comment on in Section \ref{conclu}.}:
\begin{equation}
\label{GenericScr}
    t_{scr}\sim \beta\ln(S).
\end{equation}
Here, $\beta$ is the inverse temperature of the system and $S$ can be viewed as the number of microscopic degrees of freedom in the system which take part in the fast scrambling process. The reason we do not explicitly identify $S$ with entropy of the black hole is because this is not generally true. For example, in \cite{Sekino:2008he} the scrambling time scale for very small AdS black holes\footnote{Such black holes are good approximations to asymptotically flat black holes as long as we consider processes occurring close to the horizon, fast scrambling is one such process.} is given by setting $S \sim r_H/\ell_p$, where $r_H$ is the horizon radius and $\ell_p$ is the Planck length. However, for very large AdS black holes (i.e. the ones that are thought to be dual to large $N$ CFTs) the scrambling time scale is given by setting $S \sim L/\ell_p$, where $L$ is the AdS radius. From this it follows that, for very small AdS black holes (or asymptotically flat black holes) it is reasonable to identify $S$ with the full Bekenstein-Hawking entropy of the black hole. However, for very large AdS black holes $S$ is really given by the Bekenstein-Hawking entropy of a small cell on the horizon whose characteristic length is given by the AdS radius. Indeed, this seems to be consistent with the scrambling time scale given by analyzing the behaviour of out of time order correlators \cite{Maldacena:2015waa} for large $N$ CFTs which states that the scrambling time scale is given by $t_{scr}\sim \beta \ln (N^2)$\footnote{Where we identify $(L/\ell_p)^{d-1}\sim N^2$ for large $N$ CFTs.}.

In this work, we will do a detailed analysis of the time scale set by the echo time for asymptotically $AdS_{d+1}$ black holes in various regimes. The main reason for analyzing the echo time scale for AdS black holes is because, we want to understand exactly how accurately the echo time scale can mimic the scrambling time scale.

In Section \ref{sec2}, we introduce the definition for the echo time of a spherically static black hole and define the location of the membrane in relation to the mathematical horizon. We introduce the Planck length scale by requiring the membrane is within a proper Planck length of the horizon. This enables us to expand the echo time integral as a series in the Planck length with a leading order Log term that will later be compared with the scrambling time scale. In Sections \ref{sec3} and \ref{sec4}, we explicitly calculate the echo time for different types of AdS black holes and verify the validity of the series expansion defined in Section \ref{sec2}. A central aspect of of the calculations done in Sections \ref{sec3} and \ref{sec4} is to do a detailed analysis of the $\mathcal{O}(1)$ sub-leading term in the series expansion to see how large it gets in various regimes. In Section \ref{sec5}, we compare the echo time scale and the scrambling time scale. More specifically, in Section \ref{sec5.1} we review the scrambling time scale in \cite{Sekino:2008he} and find that the scrambling time scale and echo time scale agree up to a factor of two. In Section \ref{sec:mutual} we review the results of \cite{Leichenauer:2014nxa, Brown:2018kvn} and discuss how the scrambling time scale in \cite{Leichenauer:2014nxa, Brown:2018kvn} is related to the scrambling time scale given in \cite{Sekino:2008he}. Furthermore, we review how the results of \cite{Leichenauer:2014nxa, Brown:2018kvn} suggest that there are modifications to Eq. (\ref{GenericScr}) for near extremal Reissener-Nordstrom (RN) black holes. We find that the modifications, suggested by \cite{Leichenauer:2014nxa}, to the scrambling time scale initially appears to be inconsistent with the echo time scale. We show that the discrepancy can be traced back to how one defines the smallest ``reasonable'' perturbation to a black hole. In Leichenauer's work, the smallest reasonable semi-classical perturbation is defined such that the entropy of a the black hole changes by one. We argue that this is too restrictive and propose a different definition (see Appendix \ref{BHD}) which results in an agreement between the echo and scrambling time scales in the near extremal regime. In Section \ref{discuss} we pose the question of whether echoes can exist within the framework of AdS/CFT. Based on the results of the previous sections, we give a heuristic picture of how the phenomena of echoes may be related to the phenomena of fast scrambling and what they tell us about the evolution of the Planck scale structure of the horizon. In Section \ref{conclu}, we conclude by summarizing the major findings of this paper and discuss what they imply for future studies into the connection between echoes and fast scrambling.

\section{Universal Features of Echo Time for Spherically Static Black Holes}
\label{sec2}
\subsection{Defining Echo Time}

In this section, we will introduce the exact definition of the echo time we will be using in this paper. To simplify our calculations we will restrict our discussions to spherically symmetric $d+1$-dimensional black hole metrics of the form:
\begin{equation}
\label{metric}
ds^2=-f(r)dt^2+\frac{dr^2}{f(r)}+r^2d\Omega_{d-1}^2,
\end{equation}
with $d\geq 3$. The echo time, in the geometric optics approximation is \cite{Abedi:2016hgu,Wang:2018gin}:
\begin{equation}
\label{techo}
    t_{echo}=2\int_{r_H+\delta r}^{r_t}\frac{dr}{f(r)},
\end{equation}
which is the coordinate time it takes for a radial null geodesic to go from $r_t$ to $r_H+ \delta r$ and back (hence the factor of two). Here, $r=r_H+\delta r$ is the location of the semi-reflective membrane, with $r_H$ being the location of the event horizon, i.e. $f(r_H)=0$. The upper bound of the integral, $r_t$, can be understood as a turning point of the effective potential that our perturbations are subject to. To understand exactly what this means we will consider a minimally coupled scalar field in a background defined by Eq. (\ref{metric}). In this case, the equation of motion for the scalar field can be simplified to a radial equation of the form:

\begin{equation}
\label{RadialWaveEq}
   \frac{d^2 \mathcal{R}}{dr_*^2}+\left( \omega^2-V_{\rm eff}(r) \right)\mathcal{R}=0.
\end{equation}
The details of the derivation of Eq. (\ref{RadialWaveEq}) as well as the exact form of the effective potential, $V_{\rm eff}$, is given in the Appendix \ref{EffectivePotDer}. We define $r_t$ as:

\begin{equation}
    r_t=\min\{r \colon  \omega^2-V_{\rm eff}(r)=0\}.
\end{equation}
With this definition, it is clear that the turning point depends on the frequency, $\omega$, of the scalar perturbation. In this paper we will be focusing on the echo time for ``low'' frequency perturbations\footnote{Recent studies \cite{Oshita:2018fqu,Oshita:2019sat} involving echoes has suggested that the reflection probability off the membrane for high frequency perturbations is exponential suppressed.}. Exactly what is meant by ``low'' frequency will be explained later and made more clear when we calculate the echo time in explicit examples. We shall see that, for our purposes, the exact value of $r_t$ will not be important in the ``low'' frequency regime. Finally, we will relate $\delta r$ to the Planck length, $\ell_p$, through the following integral expression:
\begin{equation}
\label{echoplanck}
    \ell_p=\int_{r_H}^{r_H+\delta r} \frac{dr}{\sqrt{f(r)}}.
\end{equation}
Physically this means that the membrane is a {\it proper} Planck length away from the horizon.

\subsection{Near Horizon Expansion of Echo Time}

Now that we have defined what the echo time is, we will expand  Eq. (\ref{techo}) in terms of $\ell_p$. To do this we will make the following assumptions on $f(r)$\footnote{All the assumptions we make are true for the black holes considered in this work.}:

\begin{enumerate}
  \item $f(r_H)=0$
  \item $f'(r_H)\neq 0$
  \item $f(r)$ is non zero and non-singular for $r>r_H$
\end{enumerate}

\begin{figure}[h!]
\centering
\includegraphics[width=90mm]{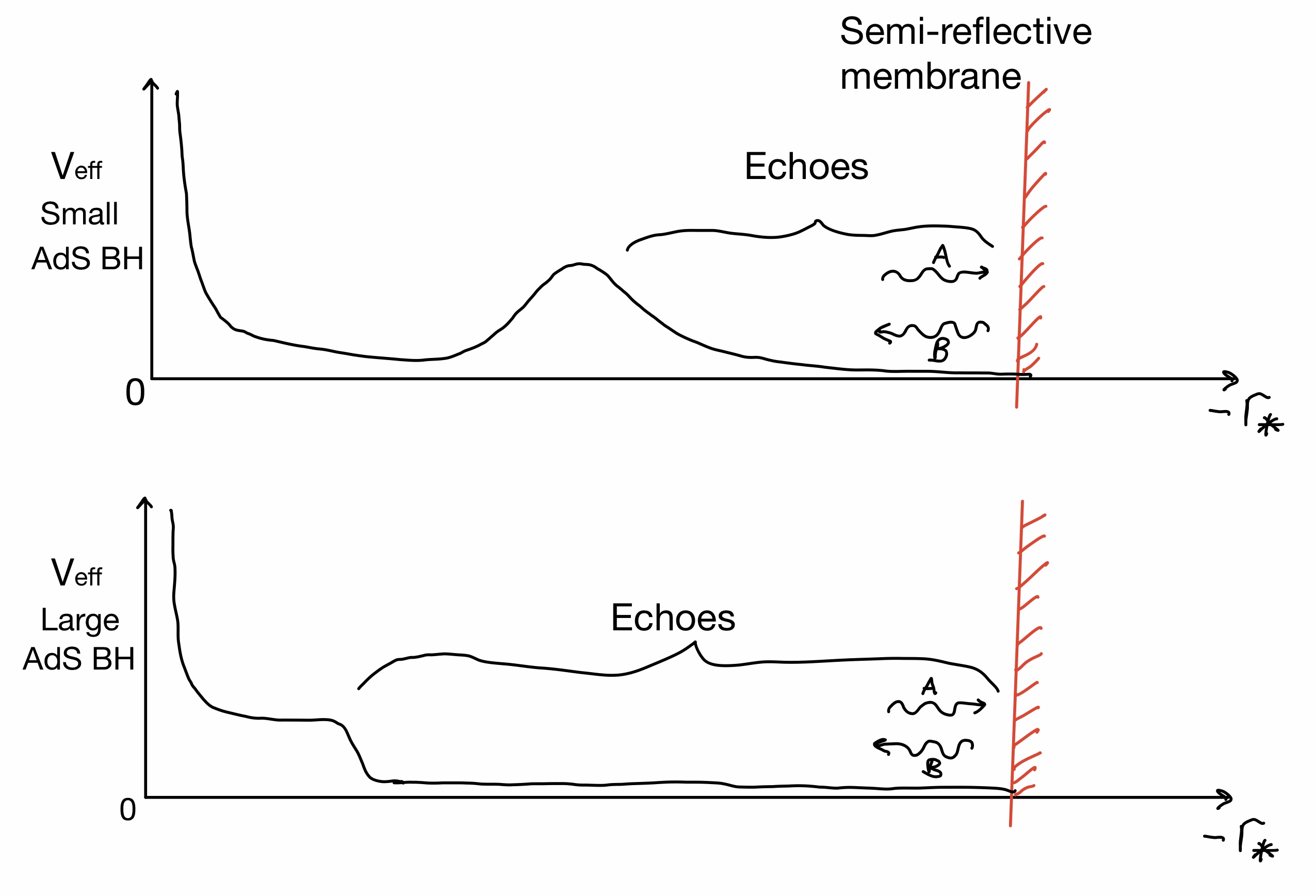}
\caption{Above is a depiction of how echoes are generated for very large ($r_H/L\gg 1$) and very small ($r_H/L\ll 1$) AdS black holes. The event horizon in these coordinates is at $r_{*}=-\infty$ and the conformal boundary is at $r_*=0$. The general solution to the massless scalar wave equation near the horizon takes the form $\psi\sim A e^{-i\omega (t+r_*)}+B e^{-i\omega (t-r_*)}$. The semi-reflective membrane, depicted by the vertical read line, allows for the partial reflection of scalar perturbations with a reflectivity of $|(B/A)e^{2i\omega r_*}|^2$. After the perturbation is partially reflected off the membrane it will head towards the conformal boundary and encounter the effective potential causing reflection back towards the membrane. The process repeats until the perturbation dissipates. For very small black holes the effective potential contains a local max before diverging near the boundary. This is in contrast to very large black holes whose effective potential continues to increase. For asymptotically flat black holes the local maximum is still present. However, there is no conformal boundary and the potential does not diverge.  \label{Veff} }
\end{figure}

With these assumptions, we will split the echo integral into two parts:
\begin{equation}
\label{techodecom}
    t_{echo}= \int_{r_H+\delta r}^{r_0}\frac{2}{f(r)}+\int_{r_0}^{r_t}\frac{2}{f(r)}.
\end{equation}
Roughly speaking $r_0$ is to be chosen such that we can do the first integral by retaining only the leading order terms in the near horizon expansion of $f(r)$. In general $r_0 \sim r_H$. This is deduced by considering the length scale set by the ratio of derivatives $|f^{(n)}(r_H)/f^{(n+1)}(r_H)|\sim r_H$. Therefore, we will write the upper limit as $r_0=Cr_H$ with $C>1$. With this we can calculate the first integral in Eq. (\ref{techodecom}):
\begin{equation}
\begin{split}
    \int_{r_H+\delta r}^{Cr_H}\frac{2dr}{f(r)}&\approx \int_{r_H+\delta r}^{Cr_H}\frac{2dr}{f'(r_H)(r-r_H)+\frac{1}{2}f''(r_H)(r-r_H)^2}\\
    &=\frac{\beta}{2\pi}\ln\left[ \frac{(C-1)r_H}{\delta r} \left( \frac{1+\frac{c_2}{c_1}\delta r}{1+(C-1)\frac{c_2}{c_1}r_H} \right)\right],\\
    \end{split}
\end{equation}
where $c_n=f^{(n)}(r_H)/n!$. It is straightforward to calculate the leading order relation between the Planck length and $\delta r$. Using Eq. (\ref{echoplanck}) we find that:

\begin{equation}
    \ell_p=\sqrt{\frac{\beta \delta r}{\pi}}\Rightarrow\delta r=\frac{\pi \ell_p^2}{\beta},
\end{equation}
where $\beta=T^{-1}=4\pi/f'(r_H)$. To simplify the final result for the leading order term we will set $C=\pi+1$\footnote{This is simply a convention  that fixes the form of the leading order Log term in the series expansion.}. Any error this introduces will be finite and of $\mathcal{O}(1)$. The $\mathcal{O}(1)$ error will be absorbed into the sub-leading terms in the Planck length expansion. With this choice of $C$, we find that:
\begin{equation}
\label{leadingorderinteg}
    \int_{r_H+\delta r}^{(\pi+1)r_H}\frac{2dr}{f(r)}\approx \frac{\beta}{2\pi}\left[ \ln\left( \frac{\beta r_H}{\ell_p^2} \right)-\ln\left( 1+\frac{f''(r_H)}{8}\beta r_H \right)+\mathcal{O}(\ell_p) \right].
\end{equation}
Therefore, in general we can write the series expansion for the echo time as:
\begin{equation}
\begin{split}
\label{techoseries}
   & t_{echo}=\frac{\beta}{2\pi}\left[ \ln\left( \frac{\beta r_H}{\ell_p^2} \right) +\chi+\mathcal{O}(\ell_p) \right],\\
   &\chi=-\ln\left( 1+\frac{f''(r_H)}{8}\beta r_H \right)+\chi_0.\\
    \end{split}
\end{equation}
In Eq. (\ref{techoseries}) $\chi_0$ is roughly given by the second integral term in Eq. (\ref{techodecom}) plus any small errors we introduce by fixing $C=\pi +1$ and doing integral in Eq. (\ref{leadingorderinteg}). Consequently, the way we defined $\chi_0$ makes it impossible to know its exact value without explicitly doing the echo integral and expanding it as a series. However, we can give a sufficient condition on it being finite. In particular, we are guaranteed that $\chi_0$ is finite as long as the second integral in Eq. (\ref{techodecom}) converges. This is guaranteed if $r_t$ is finite which brings us to a more precise definition of what is meant by a ``low'' frequency perturbation. For the black holes we will be considering the effective potential will vanish at the horizon and slowly increase. Depending on the kind of black hole, the effective potential may continue to increase (for very large AdS BH as shown in Fig. \ref{Veff}) or reach a local maximum at some point, $r_c$, (for very small AdS shown in Fig. \ref{Veff} or asymptotically flat BH). In the case where a local maximum is achieved we will only allow $r_t \leq r_c$. This will naturally place an upper bound $\Omega$ on the set of frequencies we are dealing with. We will define ``low'' frequency as $\omega<\Omega$. So we see that the low frequency criterion is needed to ensure that the size of $\chi_0$ is controlled\footnote{We intentionally did not provide a definition of low frequency for large black holes whose effective potential has no local max. This is because $\chi_0$ is always finite and does not change a great deal as we increase the turning point.}. 

However, note that even if $\chi_0$ is finite this does not imply that the entire sub-leading term $\chi$ is going to be finite. This is why we decompose $\chi$ in Eq. (\ref{techoseries}) into two pieces. The Log term will be finite far from the extremal regime, but as we approach the extremal regime the Log term will become uncontrollably large. Therefore, we should combine the the Log term in the definition of $\chi$ with the leading order Log term to get the following leading order contribution to the echo time for a near extremal BH:
\begin{equation}
\label{techoExt}
    t^{ext}_{echo}\simeq \frac{\beta}{2\pi}\left[\ln\left( \frac{8}{\ell_p^2f''_{ext}(r_H)} \right) +\mathcal{O}\left(\frac{1}{\beta r_H} \right) \right].
\end{equation}
Together, Eqs. (\ref{techoseries} - \ref{techoExt}) completely characterize the behaviour of the leading order terms in the series expansion of the echo time in various important regimes. Furthermore, we are guaranteed that sub-leading terms are either finite or suppressed by the Planck length $\ell_p$. In the next section, we will explicitly calculate the echo time for various types of black holes and show that the echo time can be arranged as a series given by Eq. (\ref{techoseries}). We will give explicit expressions for $\chi$ in these examples. In particular, we will show that $\chi$ is finite for non-extremal black holes and diverges logarithmically in $\beta$ in the near extremal regime.

\section{Echo Time For AdS Schwarzschild Black Holes}
\label{sec3}
\subsection{Overview of AdS Schwarzschild Solution}
The line element of a $d+1$-dimensional AdS Schwarzschild black hole is given by Eq. (\ref{metric}) with $f(r)$ given by:

\begin{equation}
    \begin{split}
        f(r)=1-\frac{2M}{r^{d-2}}+\frac{r^2}{L^2},
    \end{split}
\end{equation}
where $L$ is a constant called the AdS radius and $M$ is a measure of the mass of the black hole. The largest real root of $f(r)$ is the location of the event horizon and will be denoted as $r_H$. Using this fact it is useful to rewrite $f(r)$ in terms of the horizon radius to get:
\begin{equation}
\label{fAdSSch}
    f(r)=1+\frac{r^2}{L^2}-\left(\frac{r_H}{r} \right)^{d-2}\left( 1+\frac{r_H^2}{L^2} \right).
\end{equation}
We can then easily write down an expression for the temperature of the black hole:
\begin{equation}
\label{AdSSchTemp}
    T=\frac{1}{4\pi}\frac{df}{dr}\bigg|_{r=r_H}=\frac{d r^2_H+(d-2)L^2}{4\pi r_H L^2}.
\end{equation}
Analyzing the sign of $dT/dr_H$ gives us insight about the heat capacity of AdS black holes. In particular, black holes with $r_H^2/L^2<(d-2)/d$ will have a negative heat capacity and black holes with $r_H^2/L^2>(d-2)/d$ will have a positive heat capacity. The black holes with positive heat capacity are commonly referred to as large black holes and ones with negative heat capacity are referred to as small black holes.

\subsection{Echo Time in the Planar Limit}
\begin{figure}[h!]
\centering
\includegraphics[width=90mm]{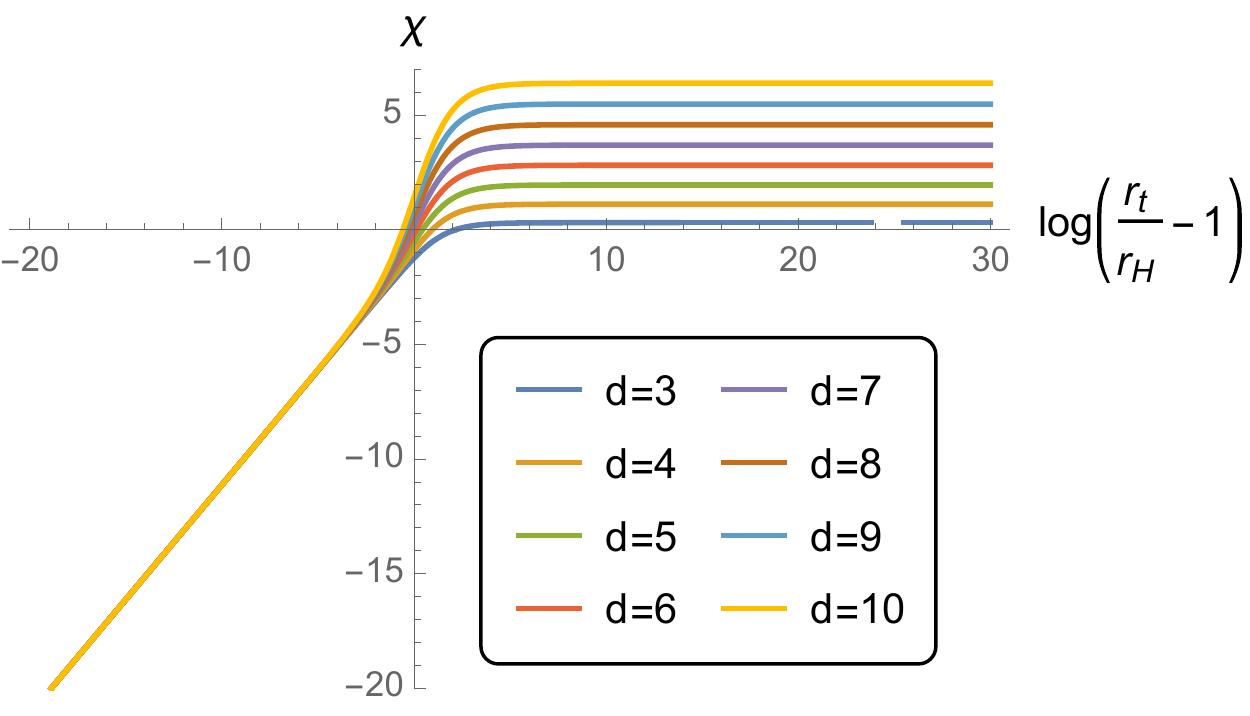}
\caption{Correction to the echo-Log, $\chi$ (defined in Eq. (\ref{techoseries})) as a function of the upper bound on echo integral (Eq. (\ref{techo})) for a $d+1$-dimensional planar (or large spherical) black hole. We see that $\chi$ asymptotes to a finite value given by Eq. (\ref{chiinftyplanarBH}) for $\chi_{\infty}$. \label{chilargeBHfig} }
\end{figure}

Since very large AdS Schwarzschild black holes at high temperature are well approximated by planar black holes it will be useful to calculate the echo time for a planar black hole. The planar black hole metric is given by Eq. (\ref{metric}) with\footnote{This not exactly correct. Technically we have to replace $d\Omega_{d-1}$ with the metric on a $d-1$ plane. Now the solutions to the scalar wave equation will be decomposed into plane waves instead of hyper-spherical  harmonics. The large angular momentum modes maps to large linear momentum modes along the horizon.}:
\begin{equation}
    f(r)=\frac{r^d-r_H^d}{L^2r^{d-2}}.
\end{equation}
The temperature is given by:
\begin{equation}
\label{PlanarTemp}
    T=\frac{dr_H}{4\pi L^2}.
\end{equation}
In this case the echo time integral can be expressed in terms of the hyper-geometric function for $d\geq 3$ and is given by:
\begin{equation}
\begin{split}
    t_{echo}&=\int_{r_H+\delta r}^{r_t}\frac{2L^2r^{d-2}}{r^d-r_H^d}\\
    &=\frac{2L^2}{r}\left[ {}_2F_1\left( 1,-\frac{1}{d},\frac{d-1}{d},\frac{r^d}{r_H^d} \right)-1 \right]\bigg\vert_{r_H+\delta r}^{r_t}\\
    &=\frac{\beta}{2\pi}\left[ \frac{dr_H}{r}{}_2F_1\left( 1,-\frac{1}{d},\frac{d-1}{d},\frac{r^d}{r_H^d}\right)-\frac{dr_H}{r} \right] \bigg\vert_{r_H+\delta r}^{r_t}. \\
    \end{split}
\end{equation}
With some work, we can write the echo time above as a series given by Eq. (\ref{techoseries}) with $\chi(r_t,r_H)$ given by:

\begin{equation}
\begin{split}
    &\chi(r_t,r_H)=\frac{dr_H}{r_t}{}_2F_1\left( 1,-\frac{1}{d},\frac{d-1}{d},\frac{r_t^d}{r_H^d} \right)+d\left( 1-\frac{r_H}{r_t} \right)-i\pi-\alpha_d\\
    &\alpha_d=\gamma+\ln(\pi d)+\psi\left( -\frac{1}{d} \right), \\
    \end{split}
\end{equation}
where $\gamma\approx 0.577$ is the Euler-Mascheroni constant and $\psi$ is the digamma function. We define $\chi_{\infty}$ as the value of $\chi$ when we take the turning point $r_t=\infty$. For the planar black hole we get a finite result:

\begin{equation}
\label{chiinftyplanarBH}
    \lim_{r_t\to \infty}\chi\left( r_t,r_H\right)=\chi_{\infty}=-\gamma-\ln(\pi d) -\psi\left( \frac{1}{d} \right).
\end{equation}
Here, $\chi_{\infty}$ represents an upper bound on the set of all possible values of $\chi$. In other words if we find that $\chi_{\infty}$ is finite, it puts a non-trivial upper bound on $\chi$ in the series expansion given by Eq. (\ref{techoseries}). In Fig. \ref{chilargeBHfig} we plot $\chi$ as a function of the ratio $r_t/r_H$ in different dimensions. We see that in general, $\chi$ is a strictly increasing function of the turning point. This makes sense because the further the turning point is the longer it takes for the echo to go from the membrane to the turning point. Furthermore, we see that for large values of the turning point $\chi$ is approaching $\chi_{\infty}$. We can ignore the divergence in the plot as $r_t\to r_H$ because we always consider our turning points to be far away from the horizon\footnote{Actually the divergence we see is necessary. The echo time should go to zero if we approach the horizon and indeed the divergence in $\chi$ will cancel with the divergence in the leading order term as we send $\ell_p\to 0$ to give an echo time of zero.}. Most importantly the plot shows that $\chi \leq \chi_{\infty}<\infty$.

Now that we have verified that $\chi$ is finite we can safely ignore it and focus our attention to the leading order term. We can use the expression for the temperature given by Eq. (\ref{PlanarTemp}) to write down the leading order contribution to the echo time:

\begin{equation}
\label{leadingorderechoplanarBH}
    t_{echo}\simeq \frac{\beta}{2\pi}\ln\left( \frac{\beta r_H}{\ell_p^2} \right)=\frac{\beta}{2\pi}\ln\left( \frac{4\pi }{d}\frac{L^2}{\ell_p^2} \right).
\end{equation}
This expression will be useful when we start comparing scrambling time to echo time for very large AdS black holes.

\subsection{Echo Time for Asymptotically Flat Schwarzschild Black Hole}
In this subsection, we will compute the echo time for asymptotically flat Schwarzschild black holes. The reason this is interesting is because, we expect the effective potential close to the horizon of a small AdS black hole to be well approximated by the effective potential of an asymptotically flat black hole. Due to this fact, we should expect the low frequency echo time for a small AdS black hole to approximately match with echo time of an asymptotically flat black hole.

\begin{figure}[h]
\centering
\includegraphics[width=90mm]{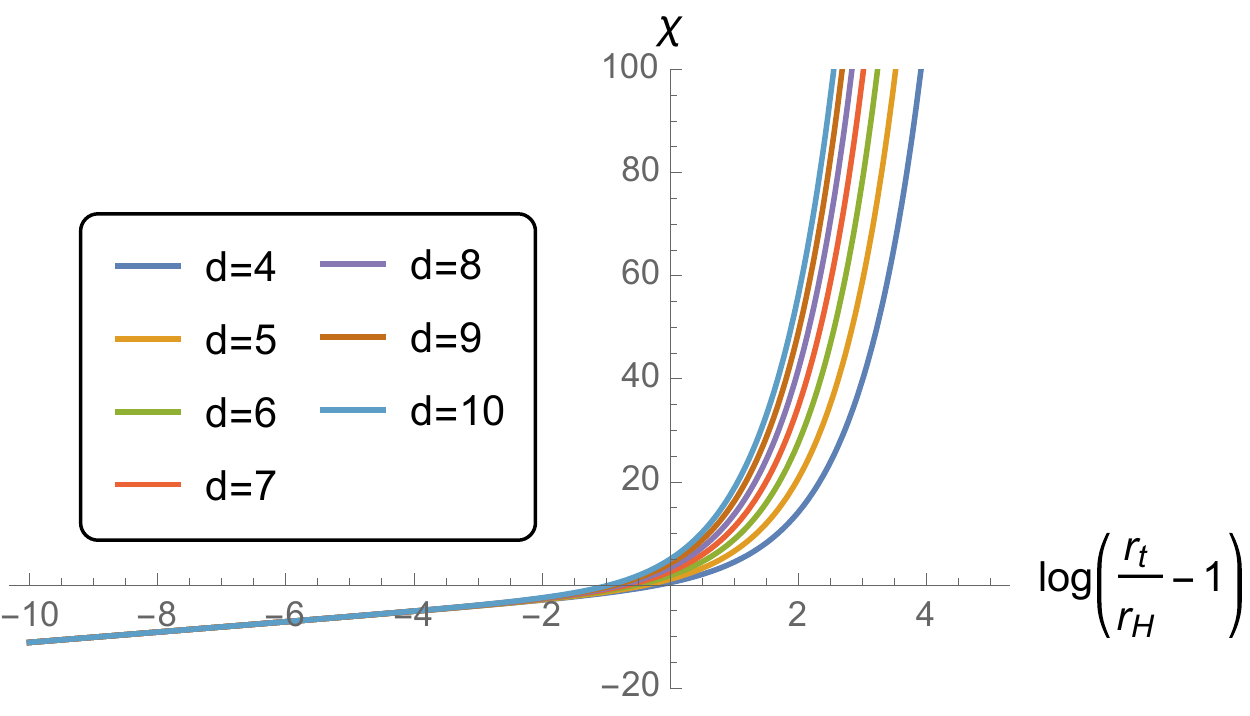}
\caption{Same as Fig. \ref{chilargeBHfig}, but for a $d+1$-dimensional asymptotically flat Schwarzschild black hole. The divergent behaviour implies we must impose a cutoff to control the how large $\chi$ becomes. The cutoff is implemented using the low frequency criterion discussed in Section \ref{sec2}. \label{chiAFBHplot}}
\end{figure}

To begin, we recall that for an asymptotically flat Schwarzschild black hole in $d+1$-dimensions $f(r)$ is given by: 
\begin{equation}
    f(r)=1-\left( \frac{r_H}{r} \right)^{d-2},
\end{equation}
and the temperature is given by:
\begin{equation}
\label{AFBHTemp}
    T=\frac{d-2}{4\pi r_H}.
\end{equation}
It follows that the echo time is given by \footnote{It should be noted that taking the limit of the above expression as $d \rightarrow 3$ is ill defined. The formula above only works for $d \geq 4$. The $d=3$ case will be calculated separately in the next subsection.}:
\begin{equation}
\begin{split}
    t_{echo}&=\frac{\beta}{2\pi}\int_{r_H+\delta r}^{r_t}\frac{(d-2)}{r_H\left(1-\left(\frac{r_H}{r}\right)^{d-2} \right)}dr\\
    &=\frac{\beta}{2\pi}\left[ (d-2)\frac{r}{r_H}{}_2F_1\left( 1,-\frac{1}{d-2},\frac{d-3}{d-2},\left(\frac{r_H}{r}\right)^{d-2} \right) \right]\Bigg\vert_{r_H+\delta r}^{r_t}. \\
    \end{split}
\end{equation}
 With some work we can eventually write the echo time in the prescribed form given by Eq. (\ref{techoseries}) with $\chi$ given by:
\begin{equation}
\begin{split}
    &\chi(r_t,r_H)=\frac{(d-2)r_t}{r_H}{}_2F_1\left( 1,\frac{1}{2-d},\frac{d-3}{d-2},\left(\frac{r_H}{r_t}\right)^{d-2} \right)-\alpha_{d-2}\\
    &\alpha_{d-2}=\gamma+\psi\left( \frac{-1}{d-2} \right)+\ln\left( \pi(d-2) \right).\\
    \end{split}
\end{equation}
Just like for the planar black hole we can plot $\chi$ as a function $r_t/r_H$ in Fig. \ref{chiAFBHplot}. 

Once again we find a strictly increasing function. However this time we find its not bounded and diverges as the turning point gets larger. The reason for this is because the point $r\to \infty$ is mapped to infinity in tortoise coordinates. To get finite results we have to restrict $r_t$ to something finite. A natural choice of the turning point is the location of the local maximum of the effective potential. It will represent the upper bound on the set of possible turning points that leave $\chi$ finite. In this case, it turns out that in the large $l$ regime we can analytically solve for the location of the local maximum. It is located at:

\begin{equation}
    r_c=\left( \frac{d}{2} \right)^{\frac{1}{d-2}}r_H.
\end{equation}

Using this we can calculate $\chi(r_c,r_H)=\chi_{max}$ and find that:

\begin{equation}
    \chi_{max}=(d-2)\left( \frac{d}{2} \right)^{\frac{1}{d-2}}{}_2F_1\left( 1,\frac{1}{2-d},\frac{d-3}{d-2},\frac{2}{d} \right)-\alpha_{d-2}.
\end{equation}
By definition we know $\chi<\chi_{max}<\infty$ and therefore finite. At this point the reader may be worried about the fact that $r_c<r_0=(\pi+1)r_H$. In Section \ref{sec2} we split the echo integral into two parts and made an implicit assumption that $r_t>r_0$ we can see that this assumption is not true here. Even so, this fact will not change the conclusion that $\chi$ is finite. However, it will change the sign of $\chi$ and make $\chi<0$. More generally, when we plot $\chi$ as a function of the turning point there will always be a set of turning points in which $\chi<0$. This will roughly correspond to when $r_t<r_0$. We say roughly because $\chi$ is not exactly given by the second integral in Eq. (\ref{techodecom}) it also contains other small errors which we discussed in Section \ref{sec2}.

Now that we have addressed the subtleties that go into making $\chi$ finite for asymptotically flat Schwarzschild black holes we can analyze what the leading order term looks like. We can use the expression for temperature given by Eq. (\ref{AFBHTemp}) to get:

\begin{equation}
\label{leadingorderechosmallBH}
    t_{echo}\simeq \frac{\beta}{2\pi}\ln\left( \frac{\beta r_H}{\ell_p^2} \right)= \frac{\beta}{2\pi}\ln\left( \frac{4\pi}{d-2}\frac{r_H^2}{\ell_p^2} \right).
\end{equation}
This expression will also be useful when we start comparing scrambling time to echo time for very small AdS black holes.

\subsection{Echo Time for 4D AdS Black Hole}
So far we have only done calculations that will give the echo time for for very large or very small AdS black holes in arbitrary dimensions. Now, we want to fix the dimension of spacetime and do the integrals without making assumptions on the size of the AdS black hole. In 4D the echo time is given by an integral of the form:
\begin{equation}
\begin{split}
     t_{echo}&=\int_{r_H+\delta r}^{r_t}\frac{2rL^2}{L^2(r-r_H)+(r^3-r_H^3)}dr\\
     &= \frac{\beta}{2\pi}\left[ \frac{2+3x_H^2}{x_H\sqrt{4+3x_H^2}}\arctan\left( \frac{2x+x_H}{\sqrt{4+3x_H^2}} \right)+\ln\left( \frac{x-x_H}{\sqrt{1+x^2+xx_H+x_H^2}} \right) \right]\Bigg \vert_{x_H+\delta x}^{x_t},\\
\end{split}
\end{equation}
where $x_H=r_H/L$, $x=r/L$, $\delta x=\delta r/L$, and $x_t=r_t/L$. We can express the result as a series expansion given by Eq. (\ref{techoseries}) with $\chi$ given by:
\begin{equation}
\begin{split}
    \chi(x_t,x_H)=&\frac{2+3x_H^2}{x_H\sqrt{4+3x_H^2}}\left[ \arctan\left( \frac{2x_t+x_H}{\sqrt{4+3x_H^2}} \right)-\arctan\left( \frac{3x_H}{\sqrt{4+3x_H^2}} \right) \right]\\
    &+\ln\left( \frac{x_t-x_H}{\pi x_H} \sqrt{\frac{1+3x_H^2}{1+x_t^2+x_tx_H+x_H^2}} \right).\\
    \end{split}
\end{equation}
We can use this result to compute $\chi$ for $d=3$ asymptotically flat black hole by taking the limit as $L\to \infty$ we find:
\begin{equation}
    \chi(r_t,r_H)=\frac{r_t}{r_H}-1+\ln\left( \frac{r_t}{r_H}-1 \right)-\ln(\pi).
\end{equation}
It is easy to see that $\chi$ is strictly increasing with the turning point and diverges with $r_t$ as expected. We can compute $\chi_{max}$ by setting $r_t=r_c=3r_H/2$ this gives:
\begin{equation}
   \chi_{max}= \frac{1}{2}-\ln\left( 2\pi \right)\approx -1.34.
\end{equation}
This completes our $d=3$ calculation for asymptotically flat black holes.

Next we calculate $\chi_{\infty}$ by taking $r_t$ to infinity this will result in the following expression:
\begin{equation}
    \chi_{\infty}(x_H)=\frac{(2+3x_H^2)\left[ \pi-2\arctan\left( \frac{3x_H}{\sqrt{4+3x_H^2}} \right) \right]}{2x_H\sqrt{4+3x_H^2}}+\ln\left( \frac{\sqrt{1+3x_H^2}}{\pi x_H} \right).
\end{equation}
\begin{figure}[h]
\label{Fig4DAdS}
\centering
\includegraphics[width=90mm]{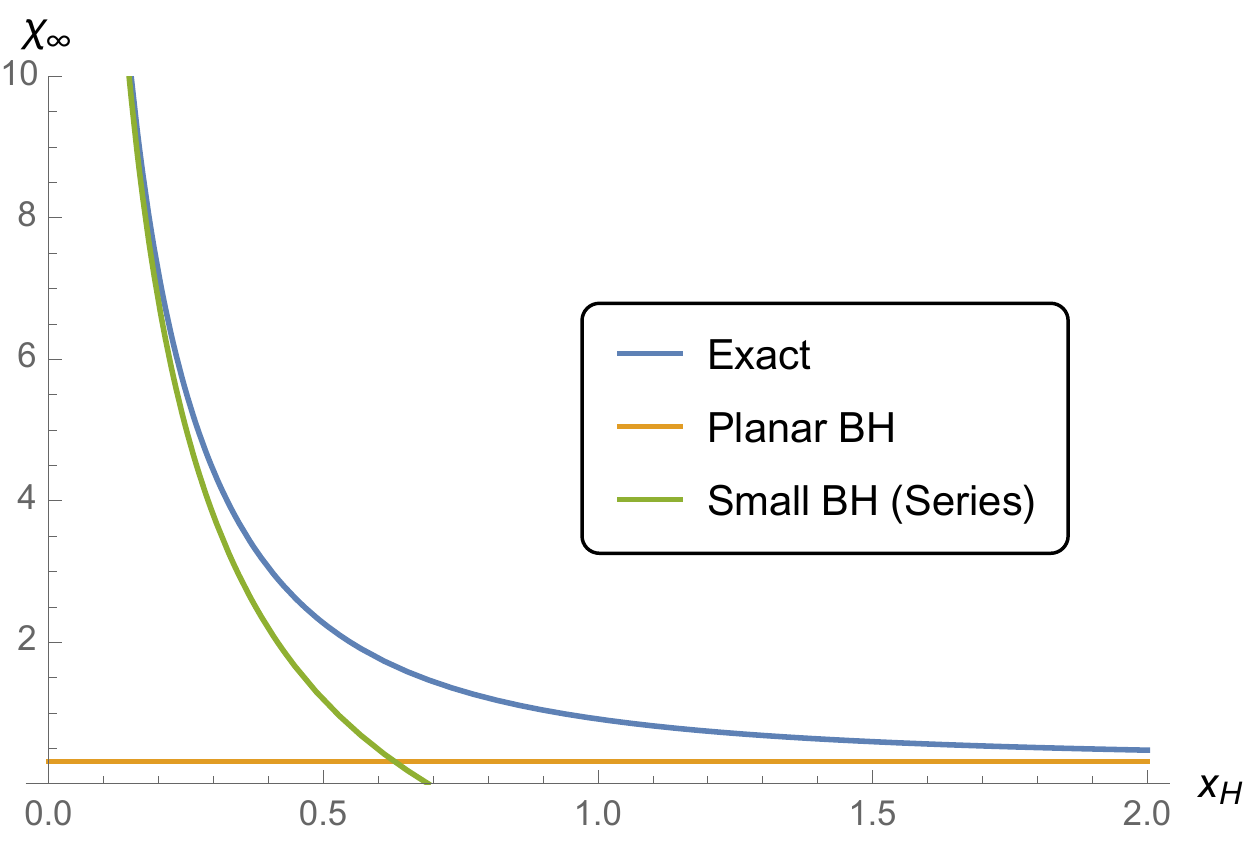}
\caption{The blue line plots $\chi_{\infty}$, the yellow line represents the planar black hole result, and the green line plots the truncated series of $\chi_{\infty}$ near $x_H=0$ given in Eq. (\ref{chiinftyser}).  \label{chiinfty4d}}
\end{figure}
We plot $\chi_{\infty}$ as a function of $x_H=r_H/L$ to get the blue line in Fig. \ref{chiinfty4d}. We see that $\chi_{\infty}$ strictly decreases and approaches the value of $\chi_{\infty}$ for the planar black hole represented by the  horizontal yellow line. The reason that $\chi_{\infty}$ is strictly decreasing is because the horizon of a larger black hole will be closer to the conformal boundary at infinity. If we analyze the behaviour of $\chi_{\infty}$ for small values of $x_H$ we will find that it diverges as $x_H \to 0$. The divergent behaviour can by deduced by analyzing the series expansion of $\chi_{\infty}$ near $x_H=0$:
\begin{equation}
\label{chiinftyser}
    \chi_{\infty}\approx \frac{\pi}{2x_H}-\ln\left( \pi x_H \right)-\frac{3}{2}+\mathcal{O}(x_H).
\end{equation}
The green line in Fig. \ref{chiinfty4d} shows that the series expansion above describes $\chi_{\infty}$ accurately for $x_H \lesssim 0.3$. This means that for very small black holes even though $\chi_{\infty}$ is finite it can become arbitrarily large for an arbitrarily small AdS black hole. However, we recall from our discussions in Section \ref{sec2} that we only want to consider low frequency modes. In such a case, the ``low'' frequency modes will encounter a local maximum in the effective potential, similar to the asymptotically flat case, before they have a chance of getting to the conformal boundary. Therefore, for low frequency modes we can ignore the fact that $\chi_{\infty}$ is unbounded for very small AdS black holes. This means that $\chi$ will always be bounded and much smaller compared to the leading order Log term in the series expansion.    

Finally, we can make the following statement about the leading order contribution to the echo time for a d+1 -dimensional AdS black hole for low frequency perturbations::    
\begin{equation}
    t_{echo}\simeq \frac{\beta}{2\pi}\ln\left( \frac{4\pi}{d(x_H^2+1)-2}\frac{r_H^2}{\ell_p^2} \right)=\begin{cases} 
      \frac{\beta}{2\pi}\left[ \ln\left(\frac{4\pi L^2}{\ell_p^2d}\right)+\mathcal{O}(1/x_H^2) \right] & x_H\gg1  \\
       \frac{\beta}{2\pi}\left[ \ln\left(\frac{4\pi r_H^2}{\ell_p^2(d-2)}\right)+\mathcal{O}(x_H^2) \right] & x_H\ll 1 \text{ and } r_t\leq r_c.
   \end{cases} 
\end{equation}
Unsurprisingly, we see that up to small corrections the leading order term for very large and very small AdS Schwarzschild black holes will match the planar black hole, Eq. (\ref{leadingorderechoplanarBH}), and Schwarzschild black hole, Eq. (\ref{leadingorderechosmallBH}), at the same temprature respectively. Similar calculations can be done in higher dimensions to verify similar results that have been explored for 4D AdS Schwarzschild black holes. Through these calculations we have explicitly checked that for non-extremal black holes $\chi$ is always finite\footnote{With the additional assumption that for very small black holes we only consider echo time for modes of sufficiently small frequency such that $r_t\leq r_c$.}.

\section{Echo Time for Reissner-Nordstrom Black Holes}
\label{sec4}
\subsection{Overview of RN Solution}
In this section, we want to understand what happens to the echo time for a Reissner-Nordstrom (RN) black hole in the near extremal regime. The RN black hole in $d+1$-spacetime dimensions is given by Eq. (\ref{metric}) with:
\begin{equation}
    f(r)=1-\frac{2M}{r^{d-2}}+\frac{Q^2}{r^{2(d-2)}}.
\end{equation}
The event horizon is given by the largest root of $f$ we can explicitly write down the roots as:
\begin{equation}
   y_{\pm}= r_{\pm}^{d-2}=M\left[ 1\pm \sqrt{1-\frac{Q^2}{M^2}} \right],
\end{equation}
where the event horizon is at $r_+$ and $r_-$ is the inner horizon. We can rewrite everything in terms of $r_{\pm}$:
\begin{equation}
    \begin{split}
        &Q^2=y_+y_-\\
        &M=\frac{1}{2}\left( y_++y_- \right)\\
        &f(r)=\frac{(r^{d-2}-r_+^{d-2})(r^{d-2}-r_-^{d-2})}{r^{2(d-2)}}.\\
    \end{split}
\end{equation}
The temperature of the black hole is given by:
\begin{equation}
\label{RNtemp}
    T=\frac{d-2}{4\pi r_+}\left[ 1-\left(\frac{r_-}{r_+}\right)^{d-2} \right].
\end{equation}
The extremal limit of the black hole occurs when $r_-=r_+$. Since we are dealing with an asymptotically flat black hole solution we should use the position of the local maximum as the turning point to get finite results. In the large $l$ regime we can find the local maximum at:

\begin{equation}
    r_c=\left[ \frac{d(r_+^{d-2}+r_-^{d-2})}{4}\left( 1+\sqrt{1-\frac{4(d-1)}{d^2}\frac{4r_-^{d-2}r_+^{d-2}}{(r_+^{d-2}+r_-^{d-2})^2}} \right) \right]^{\frac{1}{d-2}}.
\end{equation}
It can be checked that as long as $Q^2\leq M^2$ then $r_c$ is real.

\subsection{Echo Time for Non-Extremal RN Black Hole}
To calculate the echo time we need to calculate the following integral:
\begin{equation}
    t_{echo}=\int_{r_++\delta r}^{r_c}\frac{2}{\left[ 1-\left(\frac{r_+}{r}\right)^{d-2} \right]\left[ 1-\left(\frac{r_-}{r}\right)^{d-2} \right]}dr.
\end{equation}
Unfortunately, there does not appear to be a closed form for the integral unless we fix $d\geq 3$ to some particular value. As an example we can fix $d=3$. When we do this we find that:
\begin{equation}
t_{echo}^{(d=3)}=\frac{\beta}{2\pi}\left[ \frac{r(r_+-r_-)+r_+^2\ln\left(r-r_+ \right)-r_-^2\ln\left( r-r_- \right)}{r_+^2} \right]\bigg\vert_{r_++\delta r}^{r_c}.
\end{equation}  
The expression for $r_c$ when $d=3$ is:
\begin{equation}
r_c=\frac{3}{4}(r_++r_-)\left( 1+\sqrt{1-\frac{32}{9}\frac{r_+r_-}{(r_++r_-)^2}} \right).
\end{equation}
We expand the echo time in terms of the Planck length and get it into the form given by Eq. (\ref{techoseries}), where $\chi$ is given by:

\begin{equation}
\begin{split}
\chi&=\frac{1-x}{4}\left[ 3x-1+\sqrt{9+x(9x-14)} \right]\\
&+x^2\ln \left[ \frac{4(1-x)}{3-x+\sqrt{9+x(9x-14)}} \right]\\ 
&+\ln\left[ \frac{-1+3x+\sqrt{9+x(9x-14)}}{4\pi} \right],\\
\end{split}
\end{equation}
where $x=r_-/r_+$. It is not difficult to see that $\chi$ is well defined and finite away from the extremal regime. However, as we approach the extremal regime there is a divergence of the form:

\begin{equation}
\label{chid=3div}
    \chi=\ln\left( \frac{1-x}{\pi} \right)+\mathcal{O}(1-x).
\end{equation}
As expected, $\chi$ will diverge logarithmically as $x\to 1$. We will address this divergence in more detail in the next section. Assuming we are far from the extremal regime, we can write down the leading order contribution to the echo time for a d+1 -dimensional RN black hole as:
\begin{equation}
\label{techocharged}
    t_{echo}\simeq \frac{\beta}{2\pi}\ln\left( \frac{\beta r_+}{\ell_p^2} \right)=\frac{\beta}{2\pi} \ln\left( \frac{4\pi}{(d-2)(1-x^{d-2})}\frac{r_+^2}{\ell_p^2} \right).
\end{equation}
Looking at the expression above it is clear that the expression in the log is also diverging in the extremal limit.

\subsection{Echo Time for Near Extremal RN Black Hole}

In the previous section, we calculated the echo time for an RN black hole in 4D and showed that $\chi$ was divergent in the near extremal limit. If we now go towards the extremal limit and combine the result for $\chi$ given in Eq. (\ref{chid=3div}) with Eq. (\ref{techocharged}) the echo time for a 4D RN black hole is given by:

\begin{equation}
t_{echo}^{(d=3)}=\frac{\beta}{2\pi}\left[ \ln\left(\frac{4r_+^2}{\ell_p^2}\right)+\mathcal{O}(1-x)+\mathcal{O}(\ell_p) \right].
\end{equation}
We see that the divergence in $\chi$ canceled with the divergence in $\beta$ leading to a finite expression for the Log. Moreover, we can check that the leading order term in the expansion of echo time in the near extremal limit is exactly given by Eq. (\ref{techoExt}). To do this we recall that:
\begin{equation}
    f_{ext}(r)=\frac{\left( r^{d-2}-r_+^{d-2} \right)^2}{r^{2(d-2)}}\Rightarrow f_{ext}''(r_+)=\frac{2(d-2)^2}{r_+^2}.
\end{equation}
Plugging this into Eq. (\ref{techoExt}) we find:

\begin{equation}
\label{techoflatext}
    t_{echo}^{ext}\simeq \frac{\beta}{2\pi} \ln\left(\frac{4r_+^2}{(d-2)^2\ell_p^2} \right).
\end{equation}
Which correctly reproduces the leading order term in the echo time in the near extremal limit for $d=3$. One can also check this formula also works for any $d\geq 3$. This corroborates our claim that the leading order term in the echo time should look like Eq. (\ref{techoExt}) for near extremal black holes.

We can also apply Eq. (\ref{techoExt}) for very large near extremal AdS RN black holes to find:

\begin{equation}
    t_{echo}^{ext}\simeq \frac{\beta}{2\pi}\ln\left( \frac{4L^2}{d(d-1)\ell_p^2}\right).
\end{equation}
 For very small near extremal AdS RN black holes we will get the same leading order term as in the the asymptotically flat case, assuming $r_t\leq r_c$, which is given in Eq. (\ref{techoflatext}). The details of how to calculate $f_{ext}''(r_+)$ for AdS RN black holes is given in Appendix \ref{NearExtAdSRNBHAppendix}.

We can summarize the results of Section \ref{sec3} and Section \ref{sec4}  as follows. We found that the leading order term for the echo time for very large AdS black holes in both near extremal and non-extremal regimes is given by\footnote{Note that in the context of AdS/CFT the ratio $L/\ell_p$ is a measure of the effective degrees of freedom of the dual CFT state \cite{Rangamani:2016dms}. In particular, for large black holes dual to large $N$ CFTs we know $N^2\sim L^{d-1}/\ell_p^{d-1}$.}:
\begin{equation}
\label{largeechoTS}
    t_{echo}^{(Large)}\simeq \frac{\beta}{2\pi}\ln\left( \frac{L^2}{\ell_p^2} \right)\simeq \frac{2}{d-1}\frac{\beta}{2\pi}\ln(N^2).
\end{equation}
For very small or asymptotically flat black holes in both near extremal and non-extremal regimes the echo time is given by:
\begin{equation}
\label{smallechoTS}
    t_{echo}^{(Small)}\simeq \frac{\beta}{2\pi}\ln\left( \frac{r_H^2}{\ell_p^2} \right)\simeq \frac{2}{d-1}\frac{\beta}{2\pi}\ln(S_{BH}),
\end{equation}
where $S_{BH}$ is the Bekenstein-Hawking entropy of the black hole. The important fact to note here is that the echo time scale for large black holes is set by the the AdS radius and for small AdS (or asymptotically flat) black holes it is set by the horizon radius. This is consistent with the way the scrambling time scale differs for large and small AdS black holes discussed in Section \ref{sec2}.  

In the next section we will do a more detailed comparison of the time scales given by Eqs.(\ref{largeechoTS} - \ref{smallechoTS}) to the scrambling time scales given in \cite{Sekino:2008he,Leichenauer:2014nxa,Brown:2018kvn}.

\section{Echoes vs Scrambling}
\label{sec5}

\subsection{Comparison to Charge Spreading Time Scale}
\label{sec5.1}
In this section, we will compare the echo time scales given by Eqs. (\ref{largeechoTS} - \ref{smallechoTS})  with the scrambling time scale conjectured in \cite{Sekino:2008he}. We will focus on the charge spreading derivation which is done in the stretched horizon framework \cite{Susskind:1993if}. In the derivation it is assumed that the amount of time it takes for charge from a point source to spread uniformly throughout the black hole horizon can be identified with the scrambling time scale. In \cite{Sekino:2008he} the true horizon was replaced by a Rindler horizon and the charge spreading calculation was done for the Rindler horizon. With some work, which is detailed in \cite{Sekino:2008he,SusskindBookBH2005}, the following expression was derived:

\begin{equation}
    t_{sp}=\frac{\beta}{2\pi}\ln\left( \frac{\Delta x}{\ell_s}\right),
\end{equation}
where $t_{sp}$ is the Schwarzschild time it takes for the charge density to spread a distance $\Delta x$ along the horizon and $\ell_s$ is the string length\footnote{In this paper we will simply assume $\ell_s= \ell_p$ and use the two interchangeably.}. The length scale $\Delta x$, in general, cannot be identified with the horizon radius of a black hole. In particular, depending on the size of the AdS black hole, one will naturally choose either $r_H$ or $L$ length scales for $\Delta x$. In \cite{Sekino:2008he} for asymptotically flat black holes $\Delta x \sim r_H$ and for large AdS black holes $\Delta x \sim L$. Let us now discuss why these choices make sense \footnote{The argument we present is not explicitly contained in \cite{Sekino:2008he}. The authors simply identified $\Delta x$ with $r_H$ for the asymptotically flat black holes without explicitly explaining why such a choice is valid. With our argument we hope to fill in this gap.}. 

In the charge spreading calculation the true black hole horizon is replaced by a Rindler horizon. Such a replacement can only be valid within a small patch on the horizon. The size of this patch should be identified with $\Delta x$. We can estimate the length scale of the patch by calculating the Kretschmann invariant, at the horizon of the AdS black hole. To understand why the Kretschmann invariant is important one can consider Riemann normal coordinates at a point on or near the horizon. At the point of choice one is free to choose a flat metric, up to corrections second order in displacement. In other words, we are free to use use a Rindler patch. However, as we move away from this point along the horizon corrections will arise that can be written in terms of the Riemann tensor. The Riemann tensor will set an inverse length scale which should roughly be given by (the fourth root of) the Kretschmann invariant. Therefore, to suppress higher order corrections, the size of the neighborhood should be no bigger than this length scale. Now that we have an understanding of this point, let us consider the example of a 4D AdS black hole. The Kretschmann invariant is given by \cite{Hemming:2007yq}:

\begin{equation}
    R_{\mu\nu\rho\sigma}R^{\mu\nu\rho\sigma}\vert_{r=r_H}=12\left( \frac{2}{L^4}+\frac{\left(1+\frac{r_H^2}{L^2}\right)^2}{r_H^4} \right) \simeq \begin{cases} 
      \frac{36}{L^4}\left[1+\mathcal{O}(1/x_H^2)\right] & r_H \gg L  \\
      \frac{12}{r_H^4}\left[ 1+\mathcal{O}(x_H^2) \right] & r_H \ll L,
   \end{cases} 
\end{equation}
where $x_H=r_H/L$. We see that the curvature invariant sets different length scales for large and small or asymptotically flat black holes. This means that $\Delta x \sim r_H$ for small black holes and for large black holes $\Delta x\sim L$. This is consistent with the scrambling time scales suggested in \cite{Sekino:2008he}.  

Comparing to the echo time scales in Eqs. (\ref{largeechoTS} - \ref{smallechoTS}), we find agreement (up to a factor of two) between the leading order echo time scale with the charge spreading time scales for both small and large AdS black holes. Therefore, if it is reasonable to identify scrambling time scale with charge spreading then, it is also valid to identify the echo time with the scrambling time scale defined in \cite{Sekino:2008he}.

\subsection{Comparison to Mutual Information Disruption Timescale}\label{sec:mutual}
In the previous subsection, we showed that the leading order contribution to the echo time reproduces the scrambling time scale as defined by charge spreading in \cite{Sekino:2008he} (at least for non-extremal black holes). In this section, we will review how the scrambling time scale appears in Leichenauer's calculation \cite{Leichenauer:2014nxa} of mutual information disruption. After this review, we will compare with the echo time scale that we calculated.  

In \cite{Leichenauer:2014nxa} one considers a two sided RN black hole in AdS. It is known that the holographic dual to the two sided RN geometry is a charged thermofield double state of the form:
\begin{equation}
\label{TFD}
   \ket{\rm cTFD}=\frac{1}{\sqrt{Z}}\sum_{n,\sigma}e^{-\frac{\beta}{2}\left( E_n-\phi Q_\sigma \right)}\ket{n, Q_\sigma}_L\otimes\ket{n, -Q_\sigma}_R,
\end{equation}
where $\ket{n,Q_{\sigma}}_L$ and $\ket{n,Q_{\sigma}}_R$ are energy and charge eigenstates that live on the left and right conformal boundaries respectively. One can then consider two sub-regions $A$ and $B$ on the left and right field theories respectively and ask how much entanglement there is between the two sub-regions. One way of quantifying the entanglement is to calculate the mutual information which is given by:
\begin{equation}
    I(A,B)=S(A)+S(B)-S(A \cup B)\geq 0,
\end{equation}
where $S$ is the standard von Neumann entropy of the reduced density matrix of each sub-region. In general, for sufficiently large sub-regions one can show that the mutual information is non-vanishing. With these quantities in mind, one can then consider a small perturbation to the field theory on one side. This will change or disrupt the mutual information between regions $A$ and $B$. More specifically,  Leichenauer shows that the mutual information goes to zero after a time $t_*$ given by \cite{Leichenauer:2014nxa}:
\begin{equation}
\label{mutinfobdry}
    t_* \sim \frac{\beta}{2\pi} \ln\left( \frac{\Delta E}{\delta E} \right),
\end{equation}
where $\Delta E=E_{tot}-E_{ext}$, is the excess energy above the extremal energy and $\delta E$ is the energy of the perturbation\footnote{The energy above extremality of the field theory corresponds to taking the total energy $E_{tot}$ and subtracting off the energy of the field theory in the zero temperature limit, $E_{ext}$, keeping the charge fixed.}. The calculation was not directly carried out on the field theory side but instead calculated in the bulk. This was done using the Ryu-Takayanagi conjecture \cite{Ryu:2006bv,Nishioka:2009un,Rangamani:2016dms} which relates the quantities $S(A)$, $S(B)$, and $S(A\cup B)$ to the area of the extremal surfaces that extend into the bulk. The perturbation on the boundary is dual to the introduction of a shock wave that travels towards the event horizon and lengthens the wormhole connecting the two sides of the RN black hole. The disruption of mutual information occurs because the extremal surface that extends through the lengthened wormhole represents the term $S(A\cup B)$, which will also increase and cause an overall decrease in the mutual information. By considering the non-extremal regime (i.e. $\Delta E \approx E_{tot}$), it was shown that the scrambling time scale, given by Eq. (\ref{GenericScr}), is obtained by identifying $\delta E \sim E_{tot}/ S$, where $S$ is the entropy of the black hole. Using this fact it was suggested that the scrambling time scale for a near-extremal black holes should be modified to $t_{scr}\sim \beta \ln(S-S_{ext})$, where $S-S_{ext}$ is the excess entropy above the extremal black hole of the same charge.      

More recently, the same time scale has been discussed in \cite{Brown:2018kvn}. In \cite{Brown:2018kvn} the time scale derived by Leichenauer is recast completely in terms of black hole entropy rather than energy quantities on the boundary:

\begin{equation}
\label{mutinfobulk}
    t_{*}\sim \frac{\beta}{2\pi} \ln\left( \frac{S-S_{ext}}{\delta S} \right),  
\end{equation}
where $S$ is the entropy of the black hole, $S_{ext}$ is the entropy of the extremal black hole with same charge, and $\delta S$ is how much the entropy of the black hole has changed after begin perturbed. Note that, by setting $\delta S=1$ one will recover Leichenauer's modified scrambling time scale. Furthermore, we can use the first law for black hole thermodynamics and easily see that setting $\delta S=1$ corresponds to $\delta E=T_H$, where $T_H$ is the Hawking temperature of the black hole. Usually the absorption or emission of a single Hawking quantum is regarded as the smallest ``natural'' choice of perturbation to a black hole in the semi-classical regime. However, we should note that this condition might be too restrictive. For example, explicit string theory constructions of near-extremal black holes can have  $\delta S \ll 1$ (e.g., see chapter 11.3 of textbook \cite{Becker:2007zj}). Moreover, the statistical interpretation of entropy suggests that the number of microstates is given by $e^S$, implying that $\delta S_{min} \sim e^{-S}\ll 1$ (in lieu of significant degeneracies)\footnote{One way to think about this is to consider the black hole of as a collection of qubits (as is done in many considerations of scrambling in black holes) with a number of micro states equal to $W=e^S$. The smallest change in micro-states (or bits) should be larger than one. So this implies that $\delta W =e^S\delta S>1$. This in turn implies $\delta S_{min}>e^{-S}$. So even in the context of scrambling it is not necessary that $\delta S>1$. }. 

Of course, perturbations with $\delta S< 1$ will not admit to a Hawking quanta (with characteristic energy $T_H$) interpretation. However, we suggest that this may not be enough reason to disregard such perturbations in the semi-classical regime. To understand why consider the following. Suppose we have a static spherically symmetric black hole and we perturb \footnote{Assume the perturbation only changes energy and not charge or angular momentum.} it to another static spherical black hole of a different radius. Since one cannot resolve proper distances smaller than a proper Planck length it is reasonable to require that any ``measurable'' perturbation should shift the horizon by an amount larger than a proper Planck length. (We give a precise definition of what it means to shift the horizon of a black hole by a certain proper length in Appendix \ref{BHD}). For our purposes, we require that for a given $\delta R$ which corresponds to a coordinate shift in the horizon radius:

\begin{equation}
    \int_{R_H}^{R_H+\delta R}\frac{dr}{\sqrt{f(r)}} \gtrsim \ell_p.
\end{equation}
The minimal observable perturbation will saturate the constraint above and will be denoted as $\delta R_{obs}\sim \ell_p^2T_H$. Recall that the entropy of a spherically symmetric black hole in $(d+1)$ dimensions is given by:

\begin{equation}
\begin{split}
\label{EntropyOfBH}
    &S_{BH}=\frac{C_dR_H^{d-1}}{\ell_p^{d-1}}\\
    &C_d=\frac{S_{d-1}}{4}=\frac{\pi^{d/2}}{2\Gamma\left( \frac{d}{2} \right)},\\
    \end{split}
\end{equation}
where $S_{d-1}$ is the area of a $(d-1)$ unit sphere. We can take the first order variation of the entropy with respect to the horizon radius and plug in $\delta R_{obs}\sim \ell_p^2T_H$ to find: 

\begin{equation}
    \delta S_{obs} \sim \frac{R_H^{d-2}T_H}{\ell_p^{d-3}},
\end{equation}
where we dropped order one factors such as $C_d$. The expression above gives the smallest change in entropy that results in a measurable change in the horizon radius. When we deal with AdS RN black holes it is possible to have $\delta S_{obs}\ll 1$ when sufficiently close to the extremal regime (See appendix \ref{Planck_shift}). So proper Planck shifts in the near extremal regime do not have to admit to a description of perturbing by Hawking quanta with characteristic energy $T_H$. Nonetheless, you can still detect the effect of such perturbations by measuring the proper shift in the horizon. This is why it is not always necessary to discard perturbations that have $\delta S<1$ since there are alternate ways to detect a perturbation other than counting Hawking quanta. 

Going back to Eq. (\ref{mutinfobulk}) and using the choice $\delta S=\delta S_{obs}$ we will obtain time scales consistent with the echo time given by Eqs. (\ref{largeechoTS} - \ref{smallechoTS}) (see Appendix \ref{ts_extremal} for details of calculations). 

To summarize, we find that the mutual information disruption time scale defined by Eq. (\ref{mutinfobulk}) is connected to the scrambling time scale by making a choice of the smallest reasonable $\delta S$. If one chooses $\delta S=1$ one obtains Leichenauer's modified scrambling time scale for near extremal black holes. However if one instead insists that the smallest semi-classical perturbation results in observable shifts in the horizon by a proper Planck length then one will get a different time scale for scrambling consistent with the echo time. The usual choice of setting $\delta S=1$ or some other constant that is independent of any parameters specific to the black hole will always give some kind of $S-S_{ext}$ dependence inside the Log. However, as we argued these may not be the only perturbations of physical interest. One may choose perturbations that depend on parameters of the black hole. Our example of choosing perturbations that shift the horizon by a proper Planck length is one example where $\delta S$ has non-trivial $\beta$ dependence (of the form $\delta S\sim r_s/\beta$).   

More recently the scrambling time scale has also been calculated in holographic contexts that use entanglement wedge reconstruction \cite{Penington:2019npb,Almheiri:2019psf}. In particular, Pennington's work \cite{Penington:2019npb} applies to the types of black holes we have been studying in this paper. The scrambling time in his work is given by:

\begin{equation}
    t_{scr} \sim \frac{\beta}{2\pi} \ln\left( \frac{R_H}{c_{evap}\beta}\frac{R_H^{d-1}}{\ell_p^{d-1}} \right)\sim \frac{\beta}{2\pi}\ln\left( \frac{S-S_{ext}}{c_{evap}} \right).
\end{equation}
In his expressions for the scrambling time there is a parameter, $c_{evap}$, which depends on where the Hawking radiation is being extracted near the horizon. In our recent work \cite{Saraswat:2020zzf} we have shown that $c_{evap}$ will generally have non-trival dependence on $\beta$. Depending on how one chooses to extract radiation near the horizon $c_{evap}$ can have different $\beta$ dependence. This freedom/ambiguity on how we choose the $\beta$ dependence of $c_{evap}$ is similar to the freedom/ambiguity we have in choosing the $\beta$ dependence of $\delta S$ in Eq. (\ref{mutinfobulk}).

\section{Discussion: A Holographic Description of Echoes?}\label{discuss}
Thus far, we have motivated a mere mathematical relationship between that echo and scrambling time scales. In this section, we want to speculate on the physical consequences of being able to identify the scrambling time scale with the echo time scale 
in the context of AdS/CFT. 

For the sake of argument, we will assume that echoes really do exist in nature and that they owe their existence to a modification of the event horizon at Planck scales due to quantum gravity effects. Under these assumptions, it is natural to ask whether there is a holographic description of echoes within the framework of AdS/CFT. This is because the AdS/CFT correspondence claims to provide a complete description of quantum gravity in the bulk in terms of a CFT. If echoes exist in nature they should somehow also show up in the CFT description of quantum gravity. 

\begin{figure}[h]
\centering
\includegraphics[width=100mm]{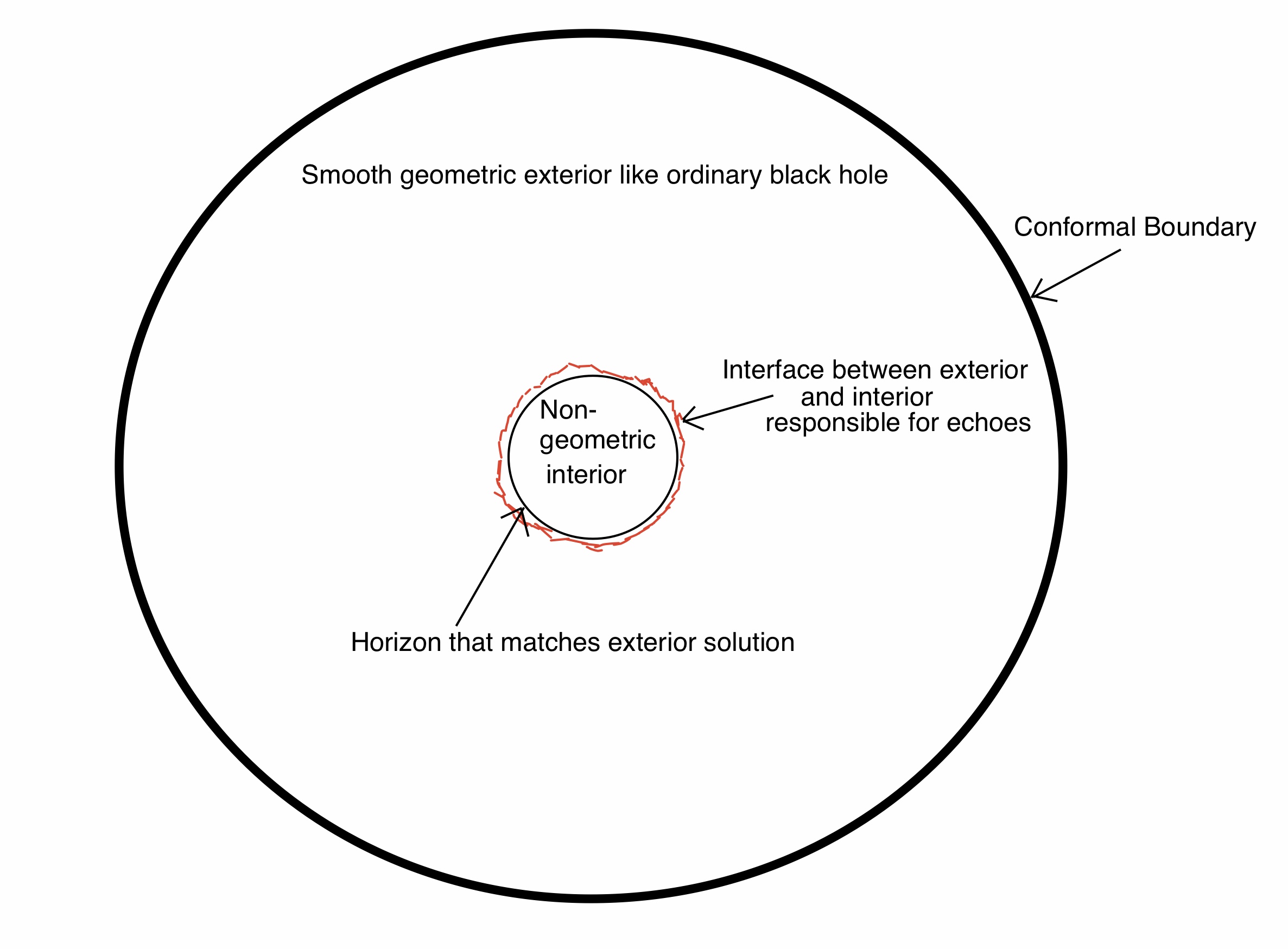}
\caption{A diagram depicting the bulk dual of a particular CFT state that exhibits echoes. The exterior far from the horizon resembles a standard black hole geometry. Within a Planck length of the horizon one expects the smooth geometrical description of spacetime to breakdown at the jagged surface colored in red. Effectively the interface between the smooth exterior and non-geometric interior generates echoes. }
\label{EchoBulk}
\end{figure}

To get an idea of how echoes might manifest themselves in a CFT calculation. It is useful to assume the existence of a state $\ket{\psi}$ which resembles a large one-sided black hole with a modified horizon. More specifically, we want the bulk dual to have a smooth geometrical description of a black hole when far away from the horizon. However, within a Planck length of the horizon the smooth geometrical picture of spacetime should break down. This is similar to the tight fuzzball proposal discussed \cite{Guo:2017jmi}. This will result in an interface between a smooth geometric exterior and a non-geometric interior as depicted in Fig. \ref{EchoBulk}. We will assume that the interface will effectively behave like the membrane that generates echoes in the bulk. We will denote this bulk spacetime as $\mathcal{M}_{\psi}$. Based on this bulk model of the CFT state $\ket{\psi}$ we will speculate how echoes in the bulk would manifest in a CFT calculation involving $\ket{\psi}$.   

\begin{figure}[h]
\centering
\includegraphics[width=100mm]{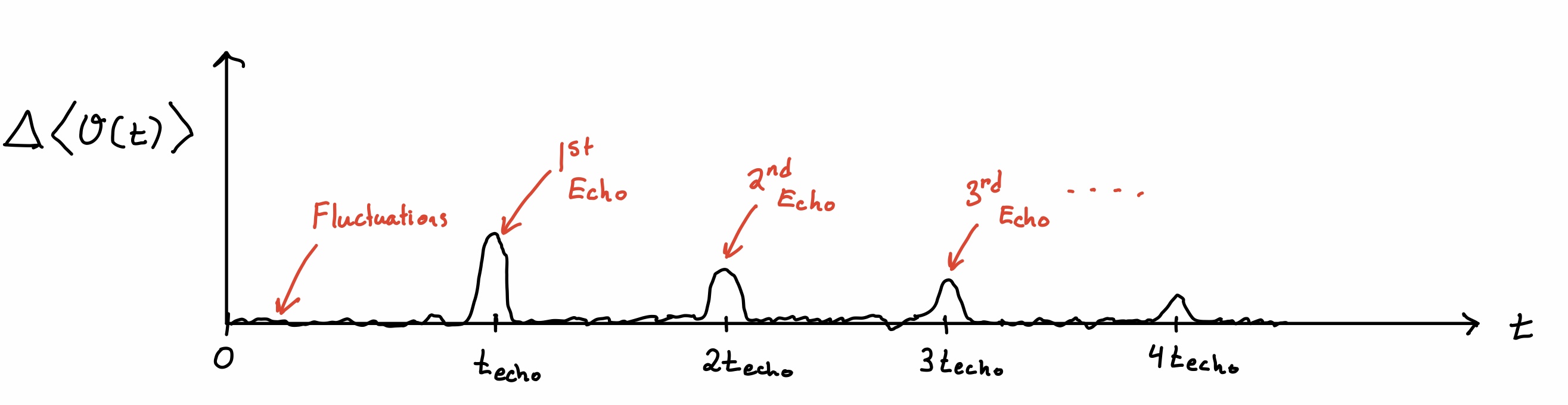}
\caption{A diagram depicting how echoes will manifest themselves in a calculation involving $\Delta \braket{\hat{O}(t-t_0)}$. Initially, the difference in the expectation value of the operator is subject to small fluctuations around zero. After one echo time scale, one would find a distinct signal above the usual fluctuations represented by the first peak. This signals the first echo in the bulk. This would reflect off the boundary and go back towards horizon and the process would repeat except subsequent echoes would gradually weaken (depicted by subsequent peaks with smaller amplitude). \label{CFTechoplot} }
\end{figure}

To start we know that if we want to ``see'' echoes we need to perturb the bulk in some way. This can be done by introducing a small perturbation near the conformal boundary in the bulk at time $t_0$. We can then consider the following quantity $\Delta\braket{\hat{O}(t)}=\bra{\psi}\hat{O}(t)\ket{\psi}-\bra{BH}\hat{O}(t)\ket{BH}$. Where $\ket{BH}$ is the CFT dual state to a black hole with a smooth horizon (i.e. same bulk as Fig. \ref{EchoBulk} without jagged red interface) and $\hat{O}(t)$ is the dual field theory operator to the perturbation in the bulk. We will refer to the smooth horizon spacetime as $\mathcal{M}_{BH}$. The time evolution of the expectation value of the operator $\hat{O}(t)$ in $\ket{\psi}$ and $\ket{BH}$ should be dual to the time evolution of the bulk perturbation around a background $\mathcal{M}_{\psi}$ and $\mathcal{M}_{BH}$ respectively. Based on the bulk geometry we should roughly expect the following behaviour:
\begin{equation}
\label{expectationvalueecho}
   \Delta\braket{\hat{O}(t)}=\bra{\psi}\hat{O}(t)\ket{\psi} -\bra{BH}\hat{O}(t)\ket{BH}\approx \begin{cases} 
      0 & 0 < t-t_0< t_{echo}  \\
      \mathcal{O}[\bra{\psi}\hat{O}(t_0)\ket{\psi}] & t-t_0 \simeq t_{echo},
   \end{cases}  
\end{equation}
To understand why this should be the case we consider what is happening in the bulk as time evolves. Initially, at $t=t_0$ the perturbation is close boundary and far from the horizon. Since $\mathcal{M}_{\psi}$ and $\mathcal{M}_{BH}$ are the same in such a region we also expect time evolution of the perturbation to be the same. However, once the perturbation gets close to the horizon it will behave differently in the two bulk spacetimes we are considering. In $\mathcal{M}_{BH}$ the perturbation will be unhindered and eventually pass through the horizon. However, in $\mathcal{M}_{\psi}$ the perturbation will encounter a reflective surface and get partially reflected back towards the conformal boundary. Information of this reflection will not arrive back at the conformal boundary until $t-t_0\simeq t_{echo}$. This is why we should expect $\Delta \braket{\hat{O}(t)}\approx 0$ when $0<t-t_0<t_{echo}$. Once the reflected perturbation hits the boundary there should be a big difference between $\bra{\psi}\hat{O}(t)\ket{\psi}$ and $\bra{BH}\hat{O}(t)\ket{BH}$ roughly of the order $\mathcal{O}[\bra{\psi}\hat{O}(t_0)\ket{\psi}]$. After this time we expect the perturbation to bounce off the conformal boundary and go back towards the horizon and repeat the same process we outlined above until the perturbation dissipates entirely. If we were to plot $\Delta\braket{\hat{O}(t-t_0)}$ we would expect a result resembling Fig. \ref{CFTechoplot}.         

Now that we have discussed how echoes in the bulk would manifest themselves in a dual CFT calculation we will discuss how we can use this picture to argue how echoes and fast scrambling can be physically related. To begin, we recall that by perturbing a black hole we can deduce the structure of the horizon by analyzing how the perturbation decays. If the decay is accompanied by echoes then it suggests the existence of a modified horizon. On the other hand, perturbing a black hole can also be regarded as the introduction of information into the black hole. As the information approaches the horizon it will become scrambled within a scrambling time scale. The process of scrambling the newly added information should gradually destroy the finely tuned entanglement between degrees of freedom close to the horizon and lead to the development of modified horizons similar to the scenarios discussed in \cite{VanRaamsdonk:2013sza,Shenker:2013pqa,Shenker:2013yza}. Eventually, the bulk geometry should evolve into configurations depicted in Fig. \ref{EchoBulk} and these types of bulk geometries would give us echoes. In other words, we believe fast scrambling to be a mechanism by which bulk geometries with smooth horizons can develop modified horizons which result in echoes. The findings of this paper which suggest that echoes and fast scrambling occur within time scales that can be reasonably identified with each other seems to be consistent with this idea.

Another interesting proposal we have on how echoes may manifest in CFT calculations is based on the work \cite{Maldacena:2015waa}. It was shown that for large $N$ CFTs, with a holographic Einstein dual, the following quantity has the following $1/N$ perturbative expansion:
\begin{equation}
\label{MaldcenaExpansion}
\begin{split}
    &F(t)=Tr[yV(0)yW(t)yV(0)yW(t)]=f_0-\frac{f_1}{N^2}e^{\frac{2\pi t}{\beta}}+\mathcal{O}(N^{-4})\\
    &y^4=\frac{1}{Z}e^{-\beta H}, \\
    \end{split}
\end{equation}
where $f_0,f_1>0$ and depend on the choice of the operators $V$ and $W$. The calculation of the sub-leading term above is done by doing a gravity calculation similar to the type of calculations done using shock wave geometries in \cite{Shenker:2013yza,Shenker:2013pqa,Leichenauer:2014nxa}. In such calculations the shock waves are perturbations to the horizon and the function $F(t)$, we suggest, should be viewed as a kind of response function which can diagnose the existence of a modified horizon. In particular, we see that the echo time for very large black holes is consistent with the scrambling time set by $\frac{\beta}{2\pi}\ln(N^2)$\footnote{Recall that $(L/\ell_p)^{d-1}\sim N^2$.} which is also when the sub-leading term in Eq. (\ref{MaldcenaExpansion}) becomes of order one. This means that the perturbative calculation after such a time scale breaks down and one needs to include higher order terms. By including all the higher order terms one might see echoes in the function $F(t)$. If this was indeed the case, it would help corroborate the claim that probing the horizon (via shock waves) will cause the horizon to develop some modified structure, which would be responsible for the echoes in $F(t)$.

\section{Conclusion}
\label{conclu}
As we already stated in the introduction of this work the existence of echoes from an experimental point of view is still tentative and controversial \cite{Abedi:2016hgu,Conklin:2017lwb,Abedi:2018npz,Westerweck:2017hus,Abedi:2018pst}. On the theoretical side, there are reasons to think that General Relativity does not tell the whole story of the nature of spacetime near the horizon of a black hole \cite{Almheiri:2012rt,Polchinski:2016hrw,Stoica:2018uli,Mathur:2009hf,Mathur:2013bra,Guo:2017jmi,Shor:2018sws,VanRaamsdonk:2013sza,Giddings:2019vvj,Giddings:2017mym}. 

In this paper, we explored the potential connection between the echo and fast scrambling time scales. We began by defining the echo time scale and explored whether it was capable of reproducing the scrambling time scale in various regimes for AdS black holes. In non-extremal regimes, we found agreement between the echo and scrambling time scales. 

For near extremal black holes,
we showed that the echo and scrambling time agree with each other for perturbations that shift the horizon by a proper Planck length. We argued in in Section \ref{sec:mutual} that the usual choice of setting $\delta S=1$ in Eq. (\ref{mutinfobulk}) as the smallest perturbation to a black hole is too restrictive. In light of this, we proposed that the smallest semi-classical perturbation should shift the horizon by a proper Planck length. The consequences of this alternate proposal is explored in depth in Appendix \ref{Planck_shift} and Appendix \ref{BHD}. Depending on the exact value of the ratios $R_{ext}/R_H$ and $R_H/L$, one will see a proper Planck shift results in different changes in the entropy (details in Appendix \ref{Planck_shift}). In general, the farther one is from the extremal regime, the larger the entropy change is for a proper Planck shift in the horizon. Furthermore, in the limit where $R_H/L$ becomes arbitrarily large, one can get arbitrarily close to an extremal black hole before $\delta S<1$.

In Section \ref{discuss}, we speculated on how echoes might manifest themselves in the context of AdS/CFT. We postulated the existence of a state $\ket{\psi}$ whose dual geometry, $\mathcal{M}_{\psi}$, resembled the bulk depicted in Fig. \ref{EchoBulk}. With this correspondence we argued that the echo time represented the amount of time it takes to determine whether a bulk geometry has a smooth or modified horizon based on the time evolution of the expectation value of some operator on the boundary. We then conjectured that the phenomena of fast scrambling and echoes are related to each other in the sense that one is a precursor for the other. More specifically we argued that fast scrambling would provide a mechanism by which black holes would develop modified horizons when perturbed. The development of modified horizons would be accompanied by echoes in the thermalization behaviour of certain CFT observables. We went further and speculated that echoes may actually be found in non-perturbative calculations of quantities similar to the ones explored in \cite{Maldacena:2015waa} given by Eq. (\ref{MaldcenaExpansion}). It would be interesting to see if it is possible to perform such non-perturbative calculations. 

As interesting as the proposals in Section \ref{discuss} are there is one major problem. The problem lies in our assumption that states that resemble black holes with modified horizons actually exist. Such an assumption is critical for the discussions in Section \ref{discuss} to be valid. In order for our arguments to be convincing one should try to explicitly find or construct a CFT state and show it exhibits echoes when perturbed. At the moment we do not have a concrete way of constructing such a state. However, it is interesting to draw upon the work of Shenker and Stanford \cite{Shenker:2013yza} which discusses the holographic dual of the thermofield double state begin perturbed by strings of operators referred to as ``thermal-scale operators.'' The bulk interpretation of such a state is that of a two sided black hole with a smooth horizon connected by a very long wormhole. Perhaps in a similar way if one acts with more generic operators on the thermofield double one might transition from a black hole with a smooth horizon to a black hole with a modified horizon. If so, such states may exhibit echoes in the way we discussed in Section \ref{discuss}.

A recent paper \cite{Oshita:2019sat}, investigating the reflectivity of modified black hole horizons, was able to show that Boltzmann reflectivity\footnote{This model assumes that the reflectivity of a modified black hole horizon depends on the frequencies of perturbations. In particular, different frequencies are weighted by a Boltzmann factor $e^{-\beta \omega}$. This means that for very high frequencies the modified horizon behaves very similar to a smooth horizon (reflectivity is approximately zero).} can be derived by considering perturbations on an $\mathbb{R}P^3$ geon. This is interesting because it provides a connection between the Boltzmann reflectivity of a modified black hole to the $\mathbb{R}P^n$ geons which have also been discussed in the context holography. In particular, work done in \cite{Guica:2014dfa} which discussed the construction $CFT_2$ states dual to the $\mathbb{R}P^2$ geon may provide ways to construct CFT states that have horizons with Boltzmann reflectivity.        

Finally, it is worth noting that not all notions of scrambling give a time scale comparable to Eq. (\ref{GenericScr}). A recent paper by Shor \cite{Shor:2018sws} suggests that in order for scrambling to occur as fast as the time scale given by Eq. (\ref{GenericScr}), via causal processes outside the stretched horizon, one needs information to leave the stretched horizon at a rate greater than what would be allowed by conventional Hawking radiation. To arrive at this conclusion, Shor used a definition of scrambling which is stronger than the definitions used in \cite{Sekino:2008he,Lashkari:2011yi,Shenker:2013yza,Leichenauer:2014nxa,Brown:2018kvn}. In particular, Shor identifies the scrambling time scale as the amount of time it takes for two unentangled hemispheres of a black hole to become maximally entangled. Naively, it seems that echoes would allow for information to escape the stretched horizon at a non-conventional rate and provide a mechanism to speed up the generation of entanglement between the two hemispheres. Therefore, it would be interesting to see if echoes can be used to speed up scrambling and make Shor's scrambling time scale consistent with Eq. (\ref{GenericScr}).

\appendix

\section{Derivation of Effective Potential for Scalar Perturbations}\label{EffectivePotDer}
In this appendix, we derive the effective potential and wave equation for a minimally coupled massless scalar field propagating a spacetime with a metric of the following form: 

\begin{equation}
ds^2=g_{\mu\nu}dx^{\mu}dx^{\nu}=-f(r)dt^2+\frac{dr^2}{f(r)}+r^2g^{\Omega}_{IJ}d\phi^Id\phi^J,
\end{equation}
where $g^{\Omega}_{IJ}$ is the metric on a $d-1$ unit sphere and $\phi^I$ are angular coordinates on the $d-1$ unit sphere. Notice that we made no assumptions of the functional form $f(r)$ so our results will work for any metric of the form given above. The equation of motion for the scalar field is a wave equation given by:

\begin{equation}
\Box \Psi=\frac{1}{\sqrt{-g}}\partial_{\mu}\left( \sqrt{-g}g^{\mu\nu}\partial_{\nu}\Psi \right)=0.
\end{equation}
Upon expansion of the sums we can write the wave equation in the form:

\begin{equation}
\Box \Psi=-\frac{1}{f(r)}\partial_t^2\Psi+\frac{1}{r^{d-1}}\partial_r\left( r^{d-1}f(r)\partial_r\Psi \right)+\frac{1}{r^2 \sqrt{g^{\Omega}}}\partial_I\left( \sqrt{g^{\Omega}}\left( g^{\Omega} \right)^{IJ}\partial_J\Psi \right)=0.
\end{equation}
We make the anzatz $\Psi=\frac{R(t,r)}{r^{\Delta}}\Phi_l(\phi^I)$, where $\Delta=\frac{d-1}{2}$ and $\Phi_l(\phi^I)$ are hyper-spherical harmonics on the unit $d-1$ sphere which obeys the eigenvalue equation:  

\begin{equation}
\label{sphhareigval}
\frac{1}{\sqrt{g^{\Omega}}}\partial_I\left( \sqrt{g^{\Omega}}\left( g^{\Omega} \right)^{IJ}\partial_J\Phi_l \right)=l(2-d-l)\Phi_l.
\end{equation}
Using the anzatz outlined above along with the eigenvalue expression for the hyper-spherical harmonics the wave equation can be written as:
\begin{equation}
-\partial_t^2R+\partial^2_{r_{*}}R-f(r)\left[ \frac{\Delta}{r}\frac{\partial f}{\partial r}+\frac{\Delta(d-2-\Delta)}{r^2}f(r)+\frac{l(l+d-2)}{r^2} \right]R=0.
\end{equation}
Where we introduced a simple change of variables in the radial coordinate $dr_{*}=\frac{dr}{f(r)}$. The resulting equation is a simple radial wave equation with an effective potential given by:

\begin{equation}
\begin{split}
&-\partial_t^2R+\partial^2_{r_{*}}R-V_{\rm eff}(r)R=0 \\
&V_{\rm eff}(r)=f(r)\left[ \frac{d-1}{2r}\frac{\partial f}{\partial r}+\frac{(d-1)(d-3)}{4r^2}f(r)+\frac{l(l+d-2)}{r^2} \right].\\
\end{split}
\end{equation} 
This gives the form of the effective potential. The angular momentum barrier occurs at a local maxima of the effective potential outside the horizon radius. In general it is not as clear that such a local maxima will exist until one specifies $f(r)$. In the large $l$ limit we can approximate the effective potential by:

\begin{equation}
    V_{\rm eff}(r)\approx \frac{l^2}{r^2}f(r).
\end{equation}
This is only valid in a finite neighborhood of the horizon but it is much easiler to analyze and find local maxima and minima of the potential in this regime. To conclude, we can plug in the Anzatz $R(t,r)=e^{-i\omega t}\mathcal{R}(r_*)$ to write down the radial equation as:

\begin{equation}
   \frac{d^2 \mathcal{R}}{dr_*^2}+\left( \omega^2-V_{\rm eff}(r) \right)\mathcal{R}=0. 
\end{equation}
The equation above makes it clear why the turning points of the effective potential depend on the frequency, $\omega$, of the scalar perturbation.

\section{Near Extremal AdS RN Black Holes}\label{NearExtAdSRNBHAppendix}
In this section we will go over the AdS RN black hole solution and its extremal regime. The AdS RN black hole has the metric given by Eq. (\ref{metric}) with:

\begin{equation}
    f(r)=1-\frac{2M}{r^{d-2}}+\frac{Q^2}{r^{2(d-2)}}+\frac{r^2}{L^2}.
\end{equation}
The horizon occurs at $r=r_+$ where $f(r_+)=0$. Using this we can rewrite $f$ in terms of the horizon radius $r_H$ and the charge $Q$:
\begin{equation}
    f(r)=\left( 1-\frac{r_+^{d-2}}{r^{d-2}} \right)\left( 1-\frac{Q^2}{r^{d-2}r_+^{d-2}} \right)+\frac{r^2}{L^2}\left( 1-\frac{r_+^d}{r^d} \right).
\end{equation}
Using this we can compute the temperature of the black hole:
\begin{equation}
\label{TempAdSRN}
    T=\frac{f'(r_+)}{4\pi} =\frac{d-2}{4\pi r_+}\left( 1-\frac{Q^2}{r_+^{2(d-2)}}+\frac{d}{d-2}\frac{r_+^2}{L^2} \right).
\end{equation}
We set the temperature equal to zero to compute the relation between $Q$ and $r_{ext}$ when the black hole is extremal. We find that:
\begin{equation}
\label{QEXT}
    Q^2=r_{ext}^{2(d-2)}\left( 1+\frac{d}{d-2}\frac{r_{ext}^2}{L^2} \right).
\end{equation}
We can plug this back into the expression for $f$ and write:

\begin{equation}
    f(r)=\left( 1-\frac{r_+^{d-2}}{r^{d-2}} \right)\left[ 1-\left( 1+\frac{d}{d-2}\frac{r_{ext}^2}{L^2} \right)\frac{r_{ext}^{2(d-2)}}{r^{d-2}r_+^{d-2}} \right]+\frac{r^2}{L^2}\left( 1-\frac{r_+^d}{r^d} \right).
\end{equation}
We get $f_{ext}(r)$ by setting $r_+=r_{ext}$. Using this we will find that:
\begin{equation}
    f''_{ext}(r_+=r_{ext})=\frac{2(d-2)^2}{r_+^2}+\frac{2d(d-1)}{L^2}.
\end{equation}
Now we can analyze what happens when $r_+\gg L$ and $r_+\ll L$:

\begin{equation}
    f''_{ext}(r_+)= \begin{cases} 
      \frac{2(d-2)^2}{r_+^2}+... & r_+\ll L  \\
      \frac{2d(d-1)}{L^2}+... & r_+\gg L.
   \end{cases} 
\end{equation}
We can use these results to compute the leading order contribution to the echo time for AdS RN black holes in the near extremal regime.

\section{Calculating $t_*$ with $\delta R=\pi \ell_p^2/\beta$}\label{ts_extremal}

To calculate $t_*$ with the choice $\delta R=\pi \ell_p^2/\beta$ it will be useful to manipulate Eq. (\ref{mutinfobulk}) as follows. Using the first law of black hole thermodynamics at constant charge we know $\delta S=\beta \delta E$. Where $\delta E$ is the energy of the perturbation.
Using the fact that the entropy of a black hole is proportional to its area ($A \sim R_H^{d-1}$) we can rewrite everything in terms of $\delta R$, $R_H$, and $R_{ext}$:

\begin{equation}
\label{tscrext}
    t_{*}\sim \frac{\beta}{2\pi} \ln\left[ \frac{R_H}{(d-1)\delta R}\left(1-\frac{R_{ext}^{d-1}}{R_H^{d-1}} \right) \right] = \begin{cases} 
      \frac{\beta}{2\pi}\left[\ln\left( \frac{R_H}{\delta R} \right) +\mathcal{O}\left(\ln\left( 1-\frac{R_{ext}^{d-1}}{R_H^{d-1}} \right)\right) \right] & R_{ext}\ll R_H  \\
      \frac{\beta}{2\pi}\left[\ln\left( \frac{R_H-R_{ext}}{\delta R} \right)+\mathcal{O}\left(1-\frac{R_{ext}}{R_H}\right)\right] & R_{ext}\approx R_H,
   \end{cases} 
\end{equation}
where $R_H$ is the radius of the black hole, $\delta R$ is the change in the radius of the black hole, and $R_{ext}$ is the radius of an extremal black hole with the same charge as the black hole we are considering. Now we set $\delta R=\pi \ell_p^2/\beta$ and then substitute this into the leading order terms in the two cases in Eq. (\ref{tscrext}) we will find:

\begin{equation}
\label{mutinfoproper}
    t_{*}\sim \begin{cases} 
      \frac{\beta}{2\pi}\ln\left( \frac{\beta R_H}{\ell_p^2} \right) & R_{ext}\ll R_H  \\
      \frac{\beta}{2\pi}\ln\left( \frac{\beta (R_H-R_{ext})}{\ell_p^2} \right) & R_{ext}\approx R_H.
   \end{cases}  
\end{equation}
From this, we can clearly see that far from the extremal limit, we reproduce the echo time scale. The second case which corresponds to a near extremal black hole requires a bit more work.

First we start with: 

\begin{equation}
   t_*^{ext}\simeq \frac{\beta}{2\pi}\ln\left( \frac{\beta (R_H-R_{ext})}{\ell_p^2} \right).
\end{equation}
Using Eqs. (\ref{TempAdSRN}-\ref{QEXT}) we can express the temperature in terms of $R_H$ and $R_{ext}$:

\begin{equation}
    T=\beta^{-1}=\frac{d-2}{4\pi R_H}\left[ \left( 1-\frac{R_{ext}^{2(d-2)}}{R_H^{2(d-2)}} \right)+\frac{d}{d-2}\frac{R_H^2}{L^2} \left( 1-\frac{R_{ext}^{2(d-1)}}{R_H^{2(d-1)}} \right) \right].
\end{equation}
Using this this we can do a series expansion for $t_*$ in the near extremal limit to get:

\begin{equation}
    t_*\simeq \frac{\beta}{2\pi}\ln\left( \frac{2\pi R_H^2}{\ell_p^2\left[ (d-2)^2+\frac{R_H^2}{L^2}d(d-1) \right]} \right)+\mathcal{O}\left(1-\frac{R_{ext}}{R_H}\right).
\end{equation} 
Using the result above we will find:

\begin{equation}
\label{tstarextprop}
    t_{*}^{ext}\sim \begin{cases}
     \frac{\beta}{2\pi}\ln\left( \frac{R_H^2}{\ell_p^2} \right) & R_H\ll L  \\
      \frac{\beta}{2\pi}\ln\left( \frac{L^2}{\ell_p^2} \right) & R_H\gg L .
    \end{cases}
\end{equation}
 Therefore, $t_*$ with the choice $\delta R=\pi \ell_p^2 T_H$, reproduces the echo time scale correctly to leading order for both large and small black holes in extremal and non-extremal regimes Eqs. (\ref{largeechoTS} - \ref{smallechoTS}).

\section{Entropy Shift corresponding to a Proper Planck Shift of the Horizon}\label{Planck_shift}
 In this appendix we study how the entropy of an AdS RN black hole changes when its horizon is shifted by a proper Planck length. We define the the physical shift in the horizon radius of a spherical black hole through the following integral expression:  

\begin{equation}
    \delta R_{phys}=\int_{R_H}^{R_H+\delta R}\frac{dr}{\sqrt{f(r)}}.
    \label{PhysicalShift}
\end{equation}
This is simply the proper length between the horizons of the unperturbed black at $R_H$ hole and the perturbed black hole at $R_H+\delta R$. Therefore, $\delta R$ is the coordinate change in the radius of the horizon which goes into the formula for calculating the entropy of a black hole. 

The the semi-classical description of spacetime as a smooth manifold is an effective description only valid on proper length scales larger than a Planck length. Due to this fact we impose the constraint, $\delta R_{phys}\gtrsim \ell_p$. Essentially, this means that the smallest possible perturbation to a black hole (which a classical observer can resolve) must shift the horizon by a proper Planck length\footnote{One may object to the way we define the shift in the horizon of a black hole on the grounds that the perturbed black hole and the unperturbed black hole are not equivalent spacetime manifolds. The integral we defined is not a good measure of how much the horizon changed because it does not account for the fact that the perturbed black hole represents a new manifold. To address this concern we show, in Appendix \ref{BHD}, that a more reasonable definition that measures the change in the horizon radius essentially gives back the same result we would get using the naive integral defined in Eq. (\ref{PhysicalShift})}.  

Now we analyze how a proper Planck shift changes the entropy content of a black hole. Using Eq. (\ref{EntropyOfBH}) and $\delta R=\pi \ell_p^2 T_H$ we can obtain the following change in the entropy of the black hole:

\begin{equation}
    \delta S=\pi C_d(d-1)R_H T_H\left( \frac{R_H}{\ell_p} \right)^{d-3}.
\end{equation}
Based on the arguments we made, the above expression represents the smallest perturbation to a black hole which results in a shift in the horizon which is classically measurable. We see that the entropy shift corresponding to a proper Planck length shift of the horizon depends on the temperature of the black hole being perturbed. In particular, there is a critical temperature below which when the change in entropy of the black hole is less than one. Setting $\delta S \geq 1$ gives us the following constraint on the temperature of the black hole:  

\begin{equation}
    T_H\geq\frac{1}{C_d(d-1)\pi R_H}\left( \frac{\ell_p}{R_H} \right)^{d-3}
\end{equation}
Now we substitute the expression for the temperature of an AdS RN black hole, given in terms of $R_{ext}$ and $R_H$ which is given by combining Eqs. (\ref{TempAdSRN} - \ref{QEXT}). We will get:

\begin{equation}
    \frac{d-2}{4\pi R_H}\left[ 1-\frac{R_{ext}^{2(d-2)}}{R_H^{2(d-2)}}+\frac{d}{d-2}\frac{R_H^2}{L^2}\left( 1-\frac{R_{ext}^{2(d-1)}}{R_H^{2(d-1)}} \right) \right] \geq\frac{1}{C_d(d-1)\pi R_H}\left( \frac{\ell_p}{R_H} \right)^{d-3}.
\end{equation}
Rearranging the terms in the inequality above gives: 

\begin{equation}
\label{NearExtCond}
    \left[ 1-\frac{R^{2(d-2)}_{ext}}{R_H^{2(d-2)}}+\frac{d}{d-2}\frac{R_H^2}{L^2}\left( 1-\frac{R^{2(d-1)}_{ext}}{R_H^{2(d-1)}} \right) \right]\geq \frac{4}{C_d(d-1)(d-2)}\left( \frac{\ell_p}{R_H} \right)^{d-3}.
\end{equation}
Where $R_H$ is the horizon radius, $R_{ext}$ is the radius if the extremal RN black hole with the same charge, and $L$ is the AdS radius. 

For an uncharged AdS Schwarzschild black hole one can set $R_{ext}=0$. In this case, it is easy to see that the constraint is satisfied whenever $R_H\gg \ell_p$. This means that for any Schwarzschild AdS black hole ($R_H\gg \ell_p$) a proper Planck shift in the horizon always changes the entropy by an amount larger than one. However, once we consider black holes sufficiently close to the extremal regime it is clear that the inequality will be violated. To understand exactly how close we need get to the extremal regime before violating the inequality. We do a series expansion of the left hand side of Eq. (\ref{NearExtCond}) near $R \simeq R_{ext}$ and find to leading order that:

\begin{equation}
    1-\frac{R_{ext}}{R_H}\gtrsim \frac{2}{C(d-1)(d-2)\left[ d-2+\frac{d(d-1)}{d-2}\frac{R_H^2}{L^2} \right]}\left( \frac{\ell_p}{R_H} \right)^{d-3}.
\end{equation}
Based on the result above, it is clear that if $d\geq 4$ we can get ``reasonably'' close to an extremal black hole (i.e. arbitrarily close in limit $\ell_p/R_H \to 0$) before a proper Planck length shift changes the entropy by an amount less than 1. 

In the case when $d=3$ we can show that Eq. (\ref{NearExtCond}) exactly takes the form of a quadratic:
\begin{equation}
\label{3dimineqy}
    -3x_H^2y^2-y+\left( 1+3x_H^2-\frac{2}{\pi} \right)\geq 0,
\end{equation}
where $y=R_{ext}^2/R_H^2$ and $x_H=R_H/L$. Taking a derivative of the left hand side of the inequality with respect to $y$ reveals that in the interval $y\in[0,1]$ the function is strictly decreasing. Furthermore, we know that the $y$-intercept of the quadratic function is positive. This means that it will become negative after it achieves its positive root. The location of the root will tell us how close we can get to the extremal regime before $\delta S<1$. Therefore, the problem simplifies to finding the positive root of the quadratic on the left hand side of the inequality. Using the quadratic formula it is easy to see that the positive zero is at:

\begin{equation}
    y=y_0=\frac{-1+\sqrt{1+12x_H^2\left( 1+3x_H^2-\frac{2}{\pi} \right)}}{6x_H^2}.
\end{equation}

\begin{figure}[h]
\centering
\includegraphics[width=100mm]{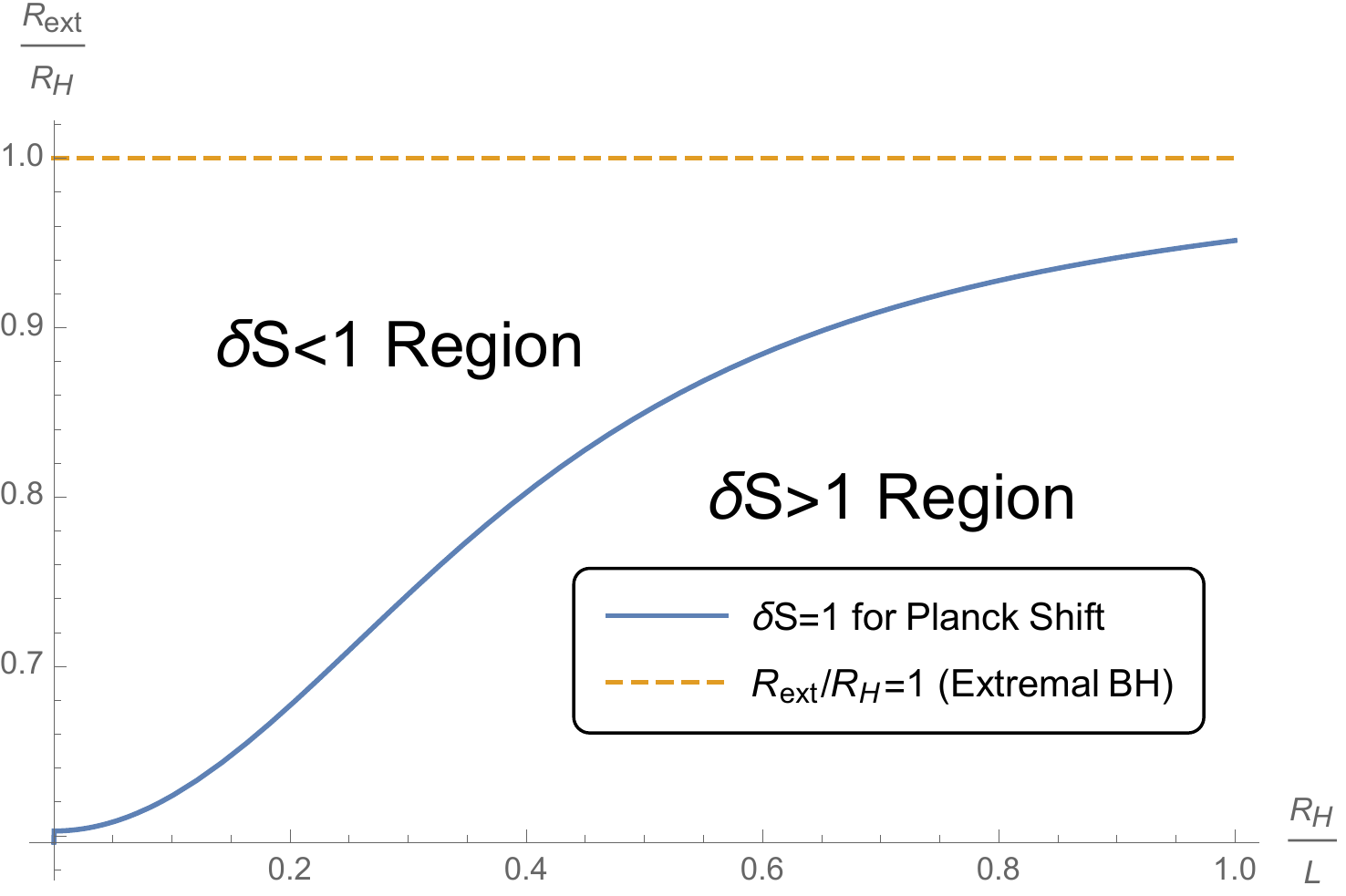}
\caption{Above is a plot of $R_{ext}/R_H$ as a function of $R_H/L$ for $d=3$ (plot of the square root of the right hand side of Eq. (\ref{d3constraintonExt})). The solid line represents the closest one can get to the extremal regime (i.e. $R_{ext}/R_H=1$ represented by the dashed line) before a proper Planck length shift of the horizon results in $\delta S<1$.}
\label{d3ExtLIMplot}
\end{figure}
Using the previous arguments it is clear that the the inequality given by Eq. (\ref{3dimineqy}) is satisfied as long as $y\in [0,y_0]$. This gives us the result in Eq. (\ref{d3constraintonExt}).

\begin{equation}
\label{d3constraintonExt}
    \frac{R_{ext}^2}{R^2}\leq \frac{-1+\sqrt{1+\frac{12R_H^2}{L^2}\left( 1+\frac{3R_H^2}{L^2}-\frac{2}{\pi} \right)}}{\frac{6R_H^2}{L^2}}.
\end{equation}
In Fig. \ref{d3ExtLIMplot} we plot the the square root of the right hand side of the inequality as a function of $R_H/L$ to get an idea of how close we can get to the extremal regime for small AdS black holes in 4D. Any black hole below the solid line whose horizon is shifted by a proper Planck length will result in $\delta S>1$. Black holes above the solid line whose horizon is shifted by a proper Planck length will have $\delta S<1$. Analyzing Fig. \ref{d3ExtLIMplot}, we see that for asymptotically flat black holes in 4D, one cannot get very close to the extremal regime (i.e. $R_{ext}/R_H \lesssim 0.6$) before a proper Planck length shift results in $\delta S<1$. However, we see that as $R_H/L$ becomes larger, one can get asymptotically closer to the extremal regime. For example, we see that once $R_H/L=1$ one can get as close as $R_{ext}/R_H\approx 0.95$. Based on these results we can conclude that that for very large AdS black holes in 4D (i.e. $R_H/L \gg 1$) we can get very close to the extremal regime before $\delta S<1$.

\section{A Semi-Classical Notion of Black Hole Distinguishability}
\label{BHD}
In this appendix we will consider a family of spherically symmetric black hole metrics, labelled by their horizon radius, which can be written in the form:

\begin{equation}
\label{FBHMetric}
    ds^2=-f_{R_H}(r)dt^2+\frac{dr^2}{f_{R_H}(r)}+r^2d\Omega_{d-1}^2.
\end{equation}
Where the subscript $R_H$ is the radial coordinate of the horizon. Now consider two black holes; one with a horizon at $R_H$ and another at $R_H+\delta R$, with $\delta R\ll R_H$. In the semi-classical regime, spacetime is described by a smooth manifold. This is an effective description which is assumed to break down on length scales smaller than a Planck length. Since there is a limit on the distances we can resolve within a spacetime it also seems reasonable to suggest that there is a limit on how well we can semi-classically distinguish two nearby black hole solutions. We propose that the following constraint should be enforced for black holes described by Eq. (\ref{FBHMetric}):

\begin{equation}
    \delta R_{phys}=\int_{R_H}^{\infty} \frac{dr}{\sqrt{f_{R_H}(r)}}-\int_{R_H+\delta R}^{\infty}\frac{dr}{\sqrt{f_{R_H+\delta R}(r)}}\gtrsim \ell_p.
\end{equation}
The constraint above describes the difference between the proper radial lengths between infinity and the horizon of two black holes. In general, the two integrals on their own will diverge. However, the difference of the integrals will converge to a finite expression. We interpret this difference as the ``proper'' change in the horizon radius and posit that in the semi-classical regime, the proper change in the radius must be larger than a Planck length. This places a constraint on the smallest possible $\delta R$ (and thereby the smallest semi-classical perturbation to a black hole). Before trying to obtain a result on the smallest $\delta R$ we will write the difference in a suggestive manner:

\begin{equation}
\label{UpgradedPlanckShift}
    \delta R_{phys}=\int_{R_H}^{R_H+\delta R}\frac{dr}{\sqrt{f_{R_H}(r)}}+\int_{R_H+\delta R}^{\infty}\left( \frac{1}{\sqrt{f_{R_H}(r)}}-\frac{1}{\sqrt{f_{R_H+\delta R}(r)}} \right).
\end{equation}
Notice that the first integral is exactly the integral given in Eq. (\ref{PhysicalShift}). We analyze the first integral in Eq. (\ref{UpgradedPlanckShift}) by expanding $f_{R_H}$ at $r=R_H$ to second order:

\begin{equation}
\begin{split}
    &\int_{R_H}^{R_H+\delta R}\frac{dr}{\sqrt{f_{R_H}(r)}}\approx \int_{R_H}^{R_H+\delta R}\frac{dr}{\sqrt{c_1(r-R_H)+c_2(r-R_H)^2}}\\
    &=\frac{2}{\sqrt{c_2}}\ln\left[ \sqrt{1+\frac{c_2}{c_1}\delta R}+\sqrt{\frac{c_2}{c_1}\delta R} \right]\\
    \end{split}
\end{equation}
Where $c_n=\frac{1}{n!}\frac{d^nf_{R_H}(R_H)}{dr^n}$, in particular $c_1= 4 \pi T_H$. We do an expansion in $\delta R$ to find (for an AdS RN BH):

\begin{equation}
\label{SeriesInegral1}
\begin{split}
    &\int_{R_H}^{R_H+\delta R}\frac{dr}{\sqrt{f_{R_H}(r)}}\approx \sqrt{\frac{\delta R}{\pi T_H}}\left[ 1-\frac{c_2\delta R}{24\pi T_H} +... \right]\\
    &c_2=\frac{1}{R_H^2}\left[ (d-2)^2+\frac{d(d-1)R_H^2}{L^2}-(2d-5)2\pi R_H T_H \right]\\
    \end{split}
\end{equation}

Now we analyze the second integral in Eq. (\ref{UpgradedPlanckShift}). To approximate the value of this integral we will again take the the example of an AdS RN black hole. We begin by expanding $f_{R_H+\delta R}(r)$ as a series in $\delta R$ to second order:

\begin{equation}
\begin{split}
    &f_{R_H+\delta R}(r)=f_{R_H}(r)+4\pi R_HT_H\left( \frac{R_H}{r} \right)^{d-2}\frac{\delta R}{R_H}\\
    &-(d-2)^2\left[ 1+\frac{d-1}{(d-2)^2}\left( \frac{dR_H^2}{L^2}-2\pi R_H T_H \right) \right]\left(\frac{R_H}{r}\right)^{d-2}\left( \frac{\delta R}{R_H} \right)^2+...\\
    \end{split}
\end{equation}
We can plug this expansion into the integrand of the second integral and expand the integrand order by order in $\delta R$. At each order in $\delta R$ we will need to evaluate integrals of the form given below: 
\begin{equation}
    I_n=\int_{R_H+\delta R}^{\infty}\frac{dr}{\left(f_{R_H}(r)\right)^{1/2+n}r^{d-2}}.
\end{equation}
These integrals clearly converge for any $n \in \mathbb{N}$ due to the fact that $f_{R_H}(r)\sim r^2/L^2$ at infinity. Since the largest contribution to the integrals will come from the lower limit of the integration, we should get a good approximation to the value of the integral by expanding $f_{R_H}(r)$ near $r=R_H$ to second order. Then we find the anti-derivative and evaluate at the lower and upper limits of integration. We then expand the result in $\delta R$ and we will find the following result:

\begin{equation}
\begin{split}
\label{SeriesIntegral2}
   &\int_{R_H+\delta R}^{\infty}\left( \frac{1}{\sqrt{f_{R_H}(r)}}-\frac{1}{\sqrt{f_{R_H+\delta R}(r)}} \right)dr \approx -\frac{5}{8}\sqrt{\frac{\delta R}{\pi T_H}}+\frac{\Gamma\left(d-\frac{3}{2}\right)}{2\Gamma(d-2)}\frac{\delta R}{\sqrt{R_HT_H}}\\
   &.-\frac{1}{64 R_H^2}\left[ \frac{5d(d-1)R_H^2}{L^2}+5(d-2)^2-2(7d-9)\pi R_H T_H \right]\left( \frac{\delta R}{\pi T_H} \right)^{3/2}+..\\
   &=\sqrt{\frac{\delta R}{\pi T_H}}\left[-\frac{5}{8}+\frac{\Gamma\left( d-\frac{3}{2}\right)}{2\Gamma(d-2)}\left(\frac{\pi \delta R}{R_H}\right)^{1/2}\right]\\
   &\sqrt{\frac{\delta R}{\pi T_H}}\left[-\frac{1}{64}\left( \frac{5d(d-1)R_H^2}{L^2}+5(d-2)^2-2(7d-9)\pi R_H T_H \right)\left( \frac{\delta R}{\pi T_HR_H^2} \right)\right]+\mathcal{O}(\delta R^2)\\
   \end{split}
\end{equation}
Once we factor out an overall factor of $\sqrt{\delta R/ (\pi T_H)}$ we can see how the series organizes itself into two parts. One part will involve terms that are multiplied by powers of $\delta R/R_H$, such terms come from the upper limit of the integral at infinity. The other part of the series involves powers of $\delta R/(T_H R_H^2)$ which come from the lower limit of the integral at $R_H+\delta R$. Once we combine the series expansions given in Eq. (\ref{SeriesInegral1}) and Eq. (\ref{SeriesIntegral2}) we will find the following terms in the expansion for $\delta R_{phys}$:

\begin{equation}
\begin{split}
    &\delta R_{phys}\approx \sqrt{\frac{\delta R}{\pi T_H}}\left[\frac{3}{8}+\frac{\Gamma\left( d-\frac{3}{2}\right)}{2\Gamma(d-2)}\left(\frac{\pi \delta R}{R_H}\right)^{1/2}\right]\\
   &+ \sqrt{\frac{\delta R}{\pi T_H}}\left[-\frac{1}{64R_H^2}\left(\frac{5d(d-1)R_H^2}{L^2}+5(d-2)^2-2(7d-9)\pi R_H T_H\right) \frac{\delta R}{\pi T_H R_H^2} \right]+...\\
     \end{split}
\end{equation}
We see that by ignoring the sub-leading terms we will have:

\begin{equation}
    \delta R_{phys}\approx \frac{3}{8}\sqrt{\frac{\delta R}{\pi T_H}} \sim \ell_p \Rightarrow \delta R \sim T_H\ell_p^2.
\end{equation}
It can be checked that if $\delta R_{phys}\sim T_H \ell_p^2$, then the sub-leading terms will be negligible as long as $\ell_p\ll min \{R_H,L\}$. 

Therefore, the leading order behaviour of $\delta R_{phys}$ is captured, up to an overall constant, by the first integral in Eq. (\ref{UpgradedPlanckShift}). This is why we can safely use the the definition given in Eq. (\ref{PhysicalShift}) to quantify the horizon shift.

\acknowledgments

We would like to thank Steve Giddings, Robb Mann, Rob Myers, Simon Ross, Rafael Sorkin, Douglas Stanford, Mark Van Raamsdonk, and Beni Yoshida  for useful discussion and comments on the manuscript.

This work was supported by the University of Waterloo, Natural Sciences and Engineering Research Council of Canada (NSERC), and the Perimeter Institute for Theoretical Physics. Research at the Perimeter Institute is supported by the Government of Canada through Industry Canada, and by the Province of Ontario through the Ministry of Research and Innovation.


\bibliographystyle{JHEP}
\bibliography{JHEPechoRef}



\end{document}